\documentclass[prd,superscriptaddress,tightenlines,nofootinbib,eqsecnum]{revtex4-2}
  
\usepackage{amssymb,amsmath,amsfonts,graphicx,mathrsfs,color,bm}
\usepackage{enumerate}
\usepackage{empheq}
\usepackage{tikz}
\usepackage{tikz-cd}
\usepackage{physics}

\usepackage{caption}

\newcommand{\Comment}[1]{{}}
\definecolor{MyDarkBlue}{rgb}{0.15,0.15,0.45}
\usepackage[linktocpage=true]{hyperref}
\usepackage{units}
\hypersetup{
colorlinks=true,
citecolor=MyDarkBlue,
linkcolor=MyDarkBlue,
urlcolor=MyDarkBlue,
pdfauthor={},
pdftitle={},
pdfsubject={}
}

\allowdisplaybreaks
\usepackage{ulem}
\newcommand{\be}{\begin{equation}}
\newcommand{\ee}{\end{equation}}
\newcommand{\sbe}{\begin{subequations}}
\newcommand{\see}{\end{subequations}}

\newcommand{\ba}{\begin{eqnarray}}
\newcommand{\ea}{\end{eqnarray}}
\newcommand{\p}{\partial}
\newcommand{\nn}{\nonumber}

\newcommand{\di}{\mathrm{i}} 
\newcommand{\de}{\mathrm{e}}
\newcommand\calO{{\mathcal{O}}}
\newcommand{\lambd}{\left(3+2\omega_0\right)}
\newcommand{\lambdprime}{3+2\omega_0}
\allowdisplaybreaks

\def\ggt{\tilde{\mathfrak{g}}}
\newcommand{\bigO}[1]{\mathcal{O}\left(\frac{1}{c^{#1}}\right)}
\newcommand{\psip}[1]{\psi_{(#1)} }
\def\switch{1 \leftrightarrow 2}

\def\W{\hat W}

\begin{document}
\title{Gravitational waves in scalar-tensor theory to one-and-a-half post-Newtonian order}

\author{Laura \textsc{Bernard}}\email{laura.bernard@obspm.fr}
\affiliation{Laboratoire Univers et Théories, Observatoire de Paris, Université PSL, Université de Paris, CNRS, F-92190 Meudon, France}

\author{Luc \textsc{Blanchet}}\email{luc.blanchet@iap.fr}
\affiliation{GReCO, Institut d’Astrophysique de Paris, UMR 7095, CNRS \& Sorbonne Université, 98bis boulevard Arago, 75014 Paris, France}

\author{David \textsc{Trestini}}\email{david.trestini@obspm.fr}
\affiliation{Laboratoire Univers et Théories, Observatoire de Paris, Université PSL, Université de Paris, CNRS, F-92190 Meudon, France}
\affiliation{GReCO, Institut d’Astrophysique de Paris, UMR 7095, CNRS \& Sorbonne Université, 98bis boulevard Arago, 75014 Paris, France}

\date{\today}

\begin{abstract}
We compute the gravitational waves generated by compact binary systems in a class of massless scalar-tensor (ST) theories to the 1.5 post-Newtonian (1.5PN) order beyond the standard quadrupole radiation in general relativity (GR). Using and adapting to ST theories the multipolar-post-Minkowskian and post-Newtonian formalisms originally defined in GR, we obtain the tail and non-linear memory terms associated with the dipole radiation in ST theory. The multipole moments and GW flux of compact binaries are derived for general orbits including the new 1.5PN contribution, and comparison is made with previous results in the literature. In the case of quasi-circular orbits, we present ready-to-use templates for the data analysis of detectors, and for the first time the scalar GW modes for comparisons with numerical relativity results.
\end{abstract}

\maketitle

\section{Introduction}\label{sec:intro}

\subsection{Context}

The multiple detections of gravitational wave events from inspiralling compact binary systems by the LIGO-Virgo observatory has allowed for a new era of gravitational wave (GW) precision astronomy~\cite{LIGOScientific:2020ibl}. They have already led to significant progress in a wide range of fields such as relativistic astrophysics, cosmology and fundamental physics. In the future, the space-based LISA observatory  coupled to the next generation of ground-based detectors, such as the Einstein Telescope, will help further improve our understanding of the Universe and the fundamental forces that govern it. However, our ability to challenge the current gravitational paradigm, based on general relativity (GR), strongly relies on the  construction of very precise and reliable banks of waveform templates for compact binary systems in various alternative theories.

In order to perform precise tests of GR with compact binary systems, one can either design some theory-dependent tests or remain agnostic on the choice and existence of a preferred theory of gravity. The former option consists in developing complete (inspiral-merger-ringdown) waveforms in a specific theory of gravity, that would complement the bank of GR templates currently being used~\cite{Khan:2015jqa,Bohe:2016gbl}. Conversely, the latter path uses general formalisms such as the parametrized post-Newtonian (PPN) formalism~\cite{Will:2018bme}, the post-Einsteinian formalism~\cite{Yunes:2009ke,Mirshekari:2011yq} and blind tests of PN parameters~\cite{Blanchet:1994ex,Mishra:2010tp}.

In this paper, we will opt for the first approach and construct gravitational waveforms within the class of scalar-tensor (ST) theories of gravity. These theories were introduced by Jordan~\cite{jordan1955schwerkraft}, Fierz~\cite{Fierz:1956zz} and Brans \& Dicke~\cite{Brans:2008zz}, and later generalized in~\cite{Nordtvedt:1970uv,Wagoner:1970vr}. Since then, they have been extensively studied, both from the theoretical and observational points of view. In particular, binary pulsar observations have already put strong constraints on the parameters entering the models, see~\cite{Damour:1992we,Fujii:2005,DeFelice:2010aj} for reviews. Due to the no-hair theorem in ST theory, valid for stationary \emph{isolated} black~holes~\cite{Hawking:1972qk,Sotiriou:2011dz}, one expects that the motion and radiation of \emph{binary} black holes are indistinguishable from those of the GR solution. To some extent, this expectation has been confirmed by numerical relativity calculations~\cite{Healy:2011ef}.
Consequently, relevant theory-dependent tests for ST theory involve binary neutron stars (or more exotic extended compact objects) as well as asymmetrical black hole-neutron star (BH-NS) binaries, such as the ones recently discovered by LIGO-Virgo~\cite{LIGOScientific:2021qlt}. 

We will focus on the inspiral phase of coalescing binary systems, for which the PN formalism, \textit{i.e.} an approximation for weak gravitational fields and small orbital velocities, can be applied. Previous results on the PN expansion in ST~theories were derived using the effective field theory~\cite{Damour:1992we,Damour:1995kt} and the Direct Integration of the Relaxed field Equations (DIRE) method~\cite{Wiseman:1992dv,Will:1996zj}. The implementation for the matched filtering analysis in the LISA detector was studied in~\cite{Scharre:2001hn}. Important previous results include the derivation of the dynamics of compact binaries at 2.5PN beyond Newtonian, \textit{i.e.}~$\sim\left(v/c\right)^5$ order~\cite{Mirshekari:2013vb}, as well as the computation of the waveform at 2PN order for the tensor modes~\cite{Lang:2013fna} and 1.5PN order for the scalar modes~\cite{Lang:2014osa} (where PN orders with respect to waveforms and fluxes are counted relatively to the quadrupolar emission, which is leading order in general relativity).

More recently, the equations of motion of compact binaries were computed to 3PN order beyond Newtonian gravity~\cite{Bernard:2018hta,Bernard:2018ivi} using the multipolar-post-Minkowskian post-Newtonian formalism (MPM-PN)~\cite{Blanchet:2013haa} coupled to a Fokker Lagrangian approach~\cite{Bernard:2015njp}. Note that at 3PN order in ST theory, the level of difficulty is similar to 4PN order in GR. The result comes with the presence of a new dipolar non-local tail term at 3PN order and the need to use a dimensional regularisation scheme to treat both the ultraviolet (UV) and infrared (IR) divergences. Finally, the 3PN results were complemented by the derivation of the scalar tidal effects to leading order in ST theories, which also arise at 3PN order due to the presence of the scalar dipole~\cite{Bernard:2019yfz}.

In the present work, we extend previous results by computing the energy flux to 1.5PN order beyond the quadrupolar radiation in GR (previously, the flux was only known to 1PN order) and we compute for the first time the scalar modes to 1.5PN order. In particular, we include the nonlinear memory effect entering at 1.5PN order associated with the dipole radiation. Recent investigations of the memory effect in ST theory can also be found in Refs.~\cite{Koyama:2020vfc,Hou:2020tnd,Seraj:2021qja}. We also treat the dipolar and quadrupolar tail effects which enter respectively at 0.5PN and 1.5PN order. Contrary to the work of Lang~\cite{Lang:2013fna,Lang:2014osa}, who uses moments named after Epstein and Wagoner (EW)~\cite{Epstein:1975}, we adopt symmetric-trace-free (STF) definitions for the multipole moments~\cite{Thorne:1980ru}. Furthermore, the formalism used in the works~\cite{Lang:2013fna,Lang:2014osa} to tackle non-linear effects is the DIRE formalism of Will and Wiseman~\cite{Will:2005sn}, which, although different, was shown to be equivalent to the present MPM-PN formalism~\cite{Blanchet:2013haa}. Thus, the present paper provides a necessary alternative derivation of the results found by Lang~\cite{Lang:2013fna,Lang:2014osa}, in addition to extending the flux by a half post-Newtonian order. 

The paper is organized as follows. In Sec~\ref{sec:STtheory}, we define the class of massless scalar-tensor theories we are interested in and derive the corresponding field equations. In Sec.~\ref{sec:MPNinST}, we adapt the multipolar post-Minkowskian formalism to the case of ST theories, focusing in particular on the memory and tail effects. Then, in Sec.~\ref{sec:compactbin}, we apply this formalism to the specific case of compact binaries, obtain the explicit expressions of the source moments and energy fluxes, and compare them with the literature. In Sec.~\ref{sec:circular}, we reduce our results to the case of circular orbits, giving the fluxes and the orbital phase evolution for dipolar-driven systems. Finally, in Sec.~\ref{sec:WF}, we compute for the first time the scalar modes, together with the previously known gravitational modes. Explicit expressions for dissipative terms in the equations of motion are also given for the first time in Appendix~\ref{app:EOM}; long results concerning the multipole moments are displayed in Appendix~\ref{app:moments} (including transformation formulae between EW and STF moments); and the very long expression of the instantaneous scalar flux is relegated to Appendix~\ref{app:fluxcoeffs}.

\subsection{Main notation and summary of parameters}
\label{sec:not}

In this section, we present the notation that will be used throughout the article. Some quantities are related to ST theories and their generalized PPN parameters, while others are linked to compact objects and binary systems.
\begin{itemize}	
\item We adopt the convention that all PN orders are relative to the standard quadrupole radiation in GR. Thus, the dominant dipole radiation enters at $-$0.5PN order in the waveform and at $-$1PN order in the energy flux. The 1.5PN order aimed at in this paper for the waveform means 2PN order beyond the leading dipolar contribution in the waveform; the 1.5PN order for the flux means 2.5PN relatively to the dipole term in the flux.
\item We use boldface letters to represent three-dimensional Euclidean vectors; we denote by $\bm{y}_{A}(t)$ the two ordinary coordinate trajectories in a harmonic coordinate system $\left\{t,\mathbf{x}\right\}$, by $\bm{v}_{A}(t)=\dd\bm{y}_{A}/\dd t$ the two ordinary velocities and by $\bm{a}_{A}(t)=\dd\bm{v}_{A}/\dd t$ the two ordinary accelerations; retarded time and advanced time are respectively denoted $u=t-r/c$ and $v=t+r/c$; the ordinary separation vector reads $\bm{n}_{12}=(\bm{y}_{1}-\bm{y}_{2})/r_{12}$, where $r_{12}=\left\vert\bm{y}_{1}-\bm{y}_{2}\right\vert$; ordinary scalar products are denoted by parentheses, \textit{e.g.} $\left(n_{12}v_{1}\right)=\bm{n}_{12}\cdot\bm{v}_{1}$, while the two masses are indicated by $m_{1}$ and $m_{2}$; the 3-dimensional Dirac function is denoted $\delta^{(3)}(\mathbf{x})$, and its value at the position $\bm{y}_A$ is written $\delta_A \equiv \delta^{(3)} (\mathbf{x} - \bm{y}_A) $.
\item Additionally, to express quantities in the center-of-mass (CM) frame, we introduce the notations $\bm{n}=\bm{n_{12}}$, $r=r_{12}$ and define the relative position $\bm{x}=\bm{y_{1}}-\bm{y_{2}}$, velocity $\bm{v}=\bm{v_{1}}-\bm{v_{2}}$, and acceleration $\bm{a}=\bm{a_{1}}-\bm{a_{2}}$; we pose $v^2=(vv)=\bm{v}\cdot\bm{v}$ and $\dot{r}=(nv)=\bm{n}\cdot\bm{v}$; the orbital frequency $\omega$ is defined by the relation $v^2=\dot{r}^2+r^2\omega^2$, and will be used when dealing with quasi-circular orbits in Sec.~\ref{sec:circular}. In the CM frame we use the total mass $m=m_1+m_2$, the reduced mass $\mu = m_1 m_2 / m$, the symmetric mass ratio $\nu= \mu/ m \in \ ]0,1/4]$ and the relative mass difference $\delta = (m_1-m_2)/m \in \ [0,1[ $. Note that the symmetric mass ratio and the relative mass difference are linked by the relation $\delta^2 = 1-4\nu$.
\item  In the case of quasi-circular orbits, we introduce $\bm{l}$ the vector normal to the orbital plane and $\boldsymbol{\lambda}$ the vector tangent to the orbit such that $\boldsymbol{\lambda}\cdot \boldsymbol{v} \ge 0 $. The vectors are oriented such that $(\mathbf{n}, \boldsymbol{\lambda}, \bm{l})$ be an orthonormal tetrad.
\item We denote $r$ the distance between the center-of-mass of the source and the observer in a harmonic coordinate system, and $n_i$ the normal vector pointing from the center-of-mass of the source towards the observer. We denote $(R, N_i)$ the same quantities in a radiative coordinate system.
\item We denote $L=i_1\cdots i_\ell$ a multi-index with $\ell$ spatial indices; $\partial_L = \partial_{i_1}\cdots\partial_{i_\ell}$, $\partial_{aL-1} = \partial_a\partial_{i_1}\cdots\partial_{i_{\ell-1}}$ and so on;  similarly,  $n_{L} = n_{i_1}\cdots n_{i_\ell}$, $n_{aL-1} = n_a n_{i_1}\cdots n_{i_{\ell-1}}$. The symmetric trace-free (STF) part is indicated using a hat or angled brackets: for instance, $\mathrm{STF}_L [\partial_L] = \hat{\partial}_L = \partial_{\langle i_1} \partial_{i_2}\cdots \partial_{ i_\ell \rangle}$, $\mathrm{STF}_L [n_L] = \hat{n}_L = n_{\langle i_1} n_{i_2}\cdots n_{ i_\ell \rangle}$ and $\mathrm{STF}_L [x_L] = \hat{x}_L = r^\ell n_{\langle i_1} n_{i_2}\cdots n_{ i_\ell \rangle}$.
\item The $n$-th time derivative of a function $F(t)$ is denoted $F^{(n)}(t) = \dd^n F / \dd t^n$, while the $n$-th time anti-derivative is $F^{(-n)}(t) = \int_{-\infty}^t \dd t_1  \int_{-\infty}^{t_1} \cdots \int_{-\infty}^{t_{n-1}} \dd t_n\, F(t_n)$. Following Eq.~(1.5) of~\cite{Blanchet:1985sp}, we assume that there exists a time $-\mathcal{T}$ in the remote past such that the fields are stationary, so that $F(t)=0$ when $t<-\mathcal{T}$ for all considered cases. This is a way to impose the \emph{no incoming radiation} condition, which isolates the system from radiation coming from sources located at past null infinity.
\item In order to later present our results, following~\cite{Bernard:2018hta}, we introduce a number of ST and post-Newtonian parameters. The ST parameters are defined based on the value $\phi_0$ of the scalar field $\phi$ at spatial infinity,  on the Brans-Dicke-like scalar function $\omega(\phi)$ and on the mass-functions $m_A(\phi)$. We pose $\varphi\equiv\phi/\phi_{0}$. The post-Newtonian parameters naturally extend and generalize the usual PPN parameters to the case of a general ST theory~\cite{Will:1972zz,Will:2018bme}. All these parameters are given and summarized in the following Table~\ref{table}.
\end{itemize}
\hspace{0.5cm}\begin{small}
\begin{center}
\begin{tabular}{|c||cc|}
	\hline
	& \multicolumn{2}{|c|}{\textbf{ST parameters}} \\[2pt]
	\hline &&\\[-10pt]
	general & \multicolumn{2}{c|}{$\omega_0=\omega(\phi_0),\qquad\omega_0'=\eval{\frac{\dd\omega}{\dd\phi}}_{\phi=\phi_0}, \qquad\omega_0''=\eval{\frac{\dd^2\omega}{\dd\phi^2}}_{\phi=\phi_0},\qquad\varphi = \frac{\phi}{\phi_{0}},\qquad\tilde{g}_{\mu\nu}=\varphi\,g_{\mu\nu},$} \\[12pt]
	& \multicolumn{2}{|c|}{$\tilde{G} = \frac{G(4+2\omega_{0})}{\phi_{0}(3+2\omega_{0})},\qquad \zeta = \frac{1}{4+2\omega_{0}},$} \\[8pt]
	& \multicolumn{2}{|c|}{$\lambda_{1} = \frac{\zeta^{2}}{(1-\zeta)}\left.\frac{\dd\omega}{\dd\varphi}\right\vert_{\varphi=1},\qquad \lambda_{2} = \frac{\zeta^{3}}{(1-\zeta)}\left.\frac{\dd^{2}\omega}{\dd\varphi^{2}}\right\vert_{\varphi=1}, \qquad \lambda_{3} = \frac{\zeta^{4}}{(1-\zeta)}\left.\frac{\dd^{3}\omega}{\dd\varphi^{3}}\right\vert_{\varphi=1}.$} \\[7pt]
	& \multicolumn{2}{|c|}{$\switch$ switches the particle's labels (note the index on the $\lambda_i$'s in not a particle label)} \\[7pt]
	\hline &&\\[-7pt]
	~sensitivities~ & \multicolumn{2}{|c|}{$s_A = \eval{\frac{\dd \ln{m_A(\phi)}}{\dd\ln{\phi}}}_{\phi=\phi_0},\qquad s_A^{(k)} = \eval{\frac{\dd^{k+1}\ln{m_A(\phi)}}{\dd(\ln{\phi})^{k+1}}}_{\phi=\phi_0},\qquad(A=1,2)$} \\[9pt]
    & \multicolumn{2}{|c|}{$s'_A = s_A^{(1)},\qquad s''_A = s_A^{(2)},\qquad s'''_A = s_A^{(3)},$} \\[5pt]
	& \multicolumn{2}{|c|}{$\mathcal{S}_+ = \frac{1-s_1 - s_2}{\sqrt{\alpha}}\,,\qquad \mathcal{S}_- = \frac{s_2 - s_1}{\sqrt{\alpha}}.$} \\[7pt]	\hline\hline 
	Order & \multicolumn{2}{|c|}{\textbf{PN parameters}} \\[2pt]
	\hline &&\\[-10pt]
	N & \multicolumn{2}{|c|}{$\alpha= 1-\zeta+\zeta\left(1-2s_{1}\right)\left(1-2s_{2}\right)$}   \\[5pt]
	\hline &&\\[-10pt]
	1PN & $\overline{\gamma} = -\frac{2\zeta}{\alpha}\left(1-2s_{1}\right)\left(1-2s_{2}\right),$ & Degeneracy \\[5pt]
	& ~~$\overline{\beta}_{1} = \frac{\zeta}{\alpha^{2}}\left(1-2s_{2}\right)^{2}\left(\lambda_{1}\left(1-2s_{1}\right)+2\zeta s'_{1}\right),$~~~~&  $\alpha(2+\overline{\gamma})=2(1-\zeta)$ \\[5pt]
	& $\overline{\beta}_{2} = \frac{\zeta}{\alpha^{2}}\left(1-2s_{1}\right)^{2}\left(\lambda_{1}\left(1-2s_{2}\right)+2\zeta s'_{2}\right),$~~~~& \\[5pt]
	&  $\overline{\beta}_+ = \frac{\overline{\beta}_1+\overline{\beta}_2}{2}, \qquad \overline{\beta}_- = \frac{\overline{\beta}_1-\overline{\beta}_2}{2}.$ &  \\[5pt]
	\hline &\\[-10pt]
	2PN & $\overline{\delta}_{1} = \frac{\zeta\left(1-\zeta\right)}{\alpha^{2}}\left(1-2s_{1}\right)^{2}\,,\qquad \overline{\delta}_{2} = \frac{\zeta\left(1-\zeta\right)}{\alpha^{2}}\left(1-2s_{2}\right)^{2},$ & Degeneracy \\[5pt]
	&  $\overline{\delta}_+ = \frac{\overline{\delta}_1+\overline{\delta}_2}{2}, \qquad \overline{\delta}_- = \frac{\overline{\delta}_1-\overline{\delta}_2}{2},$ &  $16\overline{\delta}_{1}\overline{\delta}_{2} = \overline{\gamma}^{2}(2+\overline{\gamma})^{2}$\\[5pt]
	& $~~\overline{\chi}_{1} = \frac{\zeta}{\alpha^{3}}\left(1-2s_{2}\right)^{3}\left[\left(\lambda_{2}-4\lambda_{1}^{2}+\zeta\lambda_{1}\right)\left(1-2s_{1}\right)-6\zeta\lambda_{1}s'_{1}+2\zeta^{2}s''_{1}\right],~~$  &  \\[5pt]
	& $\overline{\chi}_{2} = \frac{\zeta}{\alpha^{3}}\left(1-2s_{1}\right)^{3}\left[\left(\lambda_{2}-4\lambda_{1}^{2}+\zeta\lambda_{1}\right)\left(1-2s_{2}\right)-6\zeta\lambda_{1}s'_{2}+2\zeta^{2}s''_{2}\right],$ &  \\[5pt]
	&  $\overline{\chi}_+ = \frac{\overline{\chi}_1+\overline{\chi}_2}{2}, \qquad \overline{\chi}_- = \frac{\overline{\chi}_1-\overline{\chi}_2}{2}.$ &  \\[5pt]
\hline
\end{tabular}
\captionof{table}{Summary of parameters for the general ST theory and our notation for PN parameters. \label{table}}
\end{center}
\end{small}

\section{Massless scalar-tensor theories}\label{sec:STtheory}


We consider a generic class of scalar-tensor theories in which a single massless scalar field $\phi$ minimally couples to the metric $g_{\mu\nu}$. It is described by the action
\be\label{STactionJF}
S_{\mathrm{ST}} = \frac{c^{3}}{16\pi G} \int\dd^{4}x\,\sqrt{-g}\left[\phi R - \frac{\omega(\phi)}{\phi}g^{\alpha\beta}\p_{\alpha}\phi\p_{\beta}\phi\right] +S_{\mathrm{m}}\left(\mathfrak{m},g_{\alpha\beta}\right)\,,
\ee
where $R$ and $g$ are respectively the Ricci scalar and the determinant of the metric, $\omega$ is a function of the scalar field and $\mathfrak{m}$ stands generically for the matter fields. The action for the matter $S_{\mathrm{m}}$ is a function only of the matter fields and the metric. A major difference in scalar-tensor theories compared to GR is that, as a consequence of the breaking of the strong equivalence principle, we have to take into account the internal gravity of each body. Indeed, the scalar field determines the effective gravitational constant, which in turn affects the competition between gravitational and non-gravitational forces within the body. Thus, the value of the scalar field has an indirect influence on the size of the compact body and on its internal gravity. Here, we follow the approach pioneered by Eardley~\cite{Eardley1975} (see also~\cite{Nordtvedt:1990zz}) and we take for $S_{\mathrm{m}}$ the effective action for $N$ non-spinning point-particles with the masses $m_A(\phi)$ depending in an unspecified manner on the value of the scalar field at the location of the particles, \textit{i.e.}
\be\label{matteract}
S_{\mathrm{m}} = - c \sum_{A} \int\,m_{A}(\phi) \sqrt{-\left(g_{\alpha\beta}\right)_{A}\dd y_{A}^{\alpha}\,\dd y_{A}^{\beta}}\,,
\ee
where $y_A^\alpha$ denote the space-time positions of the particles, and $\left(g_{\alpha\beta}\right)_{A}$ is the metric evaluated at the position of particle~$A$. Thus, the matter action depends indirectly on the scalar field, and we define the sensitivities of the particles to variations in the scalar field by 
\be\label{sAk}
s_A \equiv \eval{\frac{\dd\ln{m_A(\phi)}}{\dd\ln{\phi}}}_{\phi=\phi_0}\,,\qquad\quad s_A^{(k)} \equiv \eval{\frac{\dd^{k+1}\ln{m_A(\phi)}}{\dd(\ln{\phi})^{k+1}}}_{\phi=\phi_0} \quad  \text{for } k\geqslant 2\,, 
\ee
where $\phi_0$ is the value of the scalar field at spatial infinity that is assumed to be constant in time, \textit{i.e.} we neglect the cosmological evolution. Since we expand systematically around the asymptotic value of the scalar field, we neglect scalarization~\cite{Damour:1993hw,Palenzuela:2013hsa}. The sensitivity of neutron stars depends on the mass and internal equation of state. In the weak-field limit, $s_A$ is proportional to the gravitational energy per unit mass of the body and is of order $0.2$. For stationary black holes, since all information regarding the matter which formed the black hole has disappeared behind the horizon, the mass can depend only on the Planck scale, $m_A \propto M_\text{Planck} \propto G^{-1/2} \propto \phi^{1/2}$ hence $s^\text{BH}_A=\frac{1}{2}$. In this work, we will assume that $s^\text{BH}_A=\frac{1}{2}$ for each of the black holes in a binary system and check that all our PN results will be indistinguishable from GR in the case of BBHs.

The action~\eqref{STactionJF} is usually called the ``metric'' or ``Jordan''-frame action, as the matter only couples to the Jordan or ``physical'' metric $g_{\alpha\beta}$. Then, we define the re-scaled scalar field and the conformally related metric as
\be
\varphi\equiv \frac{\phi}{\phi_{0}}\,,\qquad\qquad\tilde{g}_{\alpha\beta}\equiv \varphi\,g_{\alpha\beta}\,,
\ee
so that the physical and conformal metrics have the same asymptotic behaviour at spatial infinity. In terms of these new variables, the action~\eqref{STactionJF} can be rewritten as
\be\label{STactionEF}
S^\text{GF}_{\mathrm{ST}} = \frac{c^{3}\phi_{0}}{16\pi G} \int\dd^{4}x\,\sqrt{-\tilde{g}}\left[ \tilde{R} -\frac{1}{2}\tilde{g}_{\mu\nu}\tilde{\Gamma}^{\mu}\tilde{\Gamma}^{\nu} - \frac{3+2\omega(\phi)}{2\varphi^{2}}\tilde{g}^{\alpha\beta}\p_{\alpha}\varphi\p_{\beta}\varphi\right] +S_{\mathrm{m}}\left(\mathfrak{m},g_{\alpha\beta}\right)\,,
\ee
where we have inserted the harmonic gauge-fixing (GF) term $-\frac{1}{2}\tilde{g}_{\mu\nu}\tilde{\Gamma}^{\mu}\tilde{\Gamma}^{\nu}$ associated with the conformal metric, with $\tilde{\Gamma}^{\mu}\equiv \tilde{g}^{\rho\sigma}\tilde{\Gamma}^{\mu}_{\rho\sigma}$ and $\tilde{\Gamma}^{\mu}_{\rho\sigma}$ the Christoffel symbols of that metric, and $\tilde{R}$ the associated Ricci scalar. As the scalar field is now minimally coupled to the conformal or ``Einstein'' metric, the action~\eqref{STactionEF} is called the ``Einstein''-frame action. We will perform most of our computations in this frame and go back to the physical metric only in the end when computing observable quantities. 

Next, we define the scalar and metric perturbation variables  $\psi\equiv\varphi-1$ and $h^{\mu\nu}\equiv  \ggt^{\mu\nu} -\eta^{\mu\nu}$ where $\eta^{\mu\nu}~\equiv~\text{diag}(-1,1,1,1)$ is the Minkowski metric and $ \ggt^{\mu\nu} \equiv \sqrt{-\tilde{g}}\tilde{g}^{\mu\nu}$ is the conformal gothic metric. Then, the field equations derived from the harmonic gauge-fixed action~\eqref{STactionEF} read
\begin{subequations}\label{rEFE}
\begin{align}
& \Box_{\eta}\,h^{\mu\nu} = \frac{16\pi G}{c^{4}}\tau^{\mu\nu}\,,\\
& \Box_{\eta}\,\psi = -\frac{8\pi G}{c^{4}}\tau_{s}\,,
\end{align}
\end{subequations}
where $\Box_{\eta}$ denotes the ordinary flat space-time d'Alembertian operator, and where the source terms read
\begin{subequations}
\begin{align}\label{taumunu}
& \tau^{\mu\nu} = \frac{\varphi}{\phi_{0}} \vert g\vert T^{\mu\nu} +\frac{c^{4}}{16\pi G}\Lambda^{\mu\nu}\,,\\
\label{taus} & \tau_{s} = -\frac{\varphi}{\phi_{0}(3+2\omega)}\sqrt{-g}\left(T-2\varphi\frac{\p T}{\p \varphi}\right) -\frac{c^{4}}{8\pi G}\Lambda_s\,.
\end{align}
\end{subequations}
%
Here $T^{\mu\nu}= 2 (-g)^{-1/2}\delta S_{\mathrm{m}}/\delta g_{\mu\nu}$ is the matter stress-energy tensor, $T\equiv g_{\mu\nu}T^{\mu\nu}$ and $\p T/\p \varphi$ is defined as the partial derivative of $T(g_{\mu\nu}, \varphi)$ holding $g_{\mu\nu}$ constant. The non-linearities in the scalar source read
\be\label{lambdas} \Lambda_s = -h^{\alpha\beta}\p_{\alpha}\p_{\beta}\psi-\p_{\alpha}\psi\p_{\beta}h^{\alpha\beta} +\left(\frac{1}{\varphi}-\frac{\phi_{0}\omega'(\phi)}{3+2\omega(\phi)}\right)\tilde{\mathfrak{g}}^{\alpha\beta}\p_{\alpha}\psi\p_{\beta}\psi\,. 
\ee 
We write the tensor source as $\Lambda^{\mu\nu} = \Lambda^{\mu\nu}_{\mathrm{LL}}+\Lambda_{\mathrm{H}}^{\mu\nu}+\Lambda^{\mu\nu}_{\mathrm{GF}}+\Lambda_{\phi}^{\mu\nu}$, where $\Lambda^{\mu\nu}_{\mathrm{LL}}$ is the Landau-Lifshitz pseudo-energy tensor~\cite{Landau1971}, $\Lambda^{\mu\nu}_{\mathrm{H}}$ comes from our use of the flat version of the d'Alembertian operator in Eqs.~\eqref{rEFE}, $\Lambda^{\mu\nu}_{\mathrm{GF}}$ is due to the gauge-fixing term in the action and $\Lambda_{\phi}^{\mu\nu}$ is sourced by the scalar field. Note that $\Lambda_{\mathrm{GR}}^{\mu\nu}=\Lambda^{\mu\nu}_{\mathrm{LL}}+\Lambda_{\mathrm{H}}^{\mu\nu}+\Lambda^{\mu\nu}_{\mathrm{GF}}$ will take the same form as in GR. The expressions of these source terms are given by
\begin{subequations}\label{exprLambda}
\begin{align}\label{LambdaLL}
& \Lambda_{\mathrm{LL}}^{\alpha\beta} =\ \frac{1}{2}\tilde{\mathfrak{g}}^{\alpha\beta}\tilde{\mathfrak{g}}_{\mu\nu}\partial_{\lambda}h^{\mu\gamma}\partial_{\gamma}h^{\nu\lambda}-\tilde{\mathfrak{g}}^{\alpha\mu}\tilde{\mathfrak{g}}_{\nu\gamma}\partial_{\lambda}h^{\beta\gamma}\partial_{\mu}h^{\nu\lambda} -\tilde{\mathfrak{g}}^{\beta\mu}\tilde{\mathfrak{g}}_{\nu\gamma}\partial_{\lambda}h^{\alpha\gamma}\partial_{\mu}h^{\nu\lambda} \nn&\\
&\qquad\quad +\tilde{\mathfrak{g}}_{\mu\nu}\tilde{\mathfrak{g}}^{\lambda\gamma}\partial_{\lambda}h^{\alpha\mu}\partial_{\gamma}h^{\beta\nu} +\frac{1}{8}\left(2\tilde{\mathfrak{g}}^{\alpha\mu}\tilde{\mathfrak{g}}^{\beta\nu}-\tilde{\mathfrak{g}}^{\alpha\beta}\tilde{\mathfrak{g}}^{\mu\nu}\right)\left(2\tilde{\mathfrak{g}}_{\lambda\gamma}\tilde{\mathfrak{g}}_{\tau\pi}-\tilde{\mathfrak{g}}_{\gamma\tau}\tilde{\mathfrak{g}}_{\lambda\pi}\right)\partial_{\mu}h^{\lambda\pi}\partial_{\nu}h^{\gamma\tau} \,, \\
\label{LambdaH}
&\Lambda_{\mathrm{H}}^{\alpha\beta} = -h^{\mu\nu}\partial_{\mu}\partial_{\nu}h^{\alpha\beta} +\partial_{\mu}h^{\alpha\nu}\partial_{\nu}h^{\beta\mu} \,,\\
\label{Lambdagf}
& \Lambda_{\mathrm{GF}}^{\alpha\beta} = - \partial_{\lambda}h^{\lambda\alpha} \partial_{\sigma}h^{\sigma\beta} - \partial_{\lambda}h^{\lambda\rho}\partial_{\rho}h^{\alpha\beta} -\frac{1}{2}\tilde{\mathfrak{g}}^{\alpha\beta}\tilde{\mathfrak{g}}_{\rho\sigma} \partial_{\lambda}h^{\lambda\rho} \partial_{\gamma}h^{\gamma\sigma} +2\tilde{\mathfrak{g}}_{\rho\sigma}\tilde{\mathfrak{g}}^{\lambda(\alpha} \partial_{\lambda}h^{\beta)\rho}\partial_{\gamma}h^{\gamma\sigma} \,,\\
\label{Lambdas}
&\Lambda_{\phi}^{\mu\nu} = \frac{3+2\omega(\phi)}{\varphi^2}\left(\tilde{\mathfrak{g}}^{\mu\alpha}\tilde{\mathfrak{g}}^{\nu\beta} -\frac{1}{2}\tilde{\mathfrak{g}}^{\mu\nu}\tilde{\mathfrak{g}}^{\alpha\beta}\right)\p_{\alpha}\psi\p_{\beta}\psi\,.
\end{align}
\end{subequations}
Note that the gauge-fixing term~\eqref{Lambdagf} contains the harmonicities $\p_{\nu}h^{\mu\nu}$ which are not zero off-shell, \textit{i.e.} when the accelerations are not replaced by the equations of motion. However, this term will ensure that, on-shell, our results are in harmonic coordinates.

\section{The Multipolar post-Minkowskian formalism in scalar-tensor theories}\label{sec:MPNinST}


\subsection{The scalar-tensor multipole moments}

In this section, we solve the ST vacuum field equations, $\Box h^{\mu\nu} = \Lambda^{\mu\nu}$ and $\Box \psi = \Lambda_s$, in the exterior region of the isolated matter system by means of a multipolar decomposition (indicated by $\mathcal{M}$) conjointly with a non-linear post-Minkowskian expansion~\cite{Blanchet:1985sp}. Thus, the solution is written as
\begin{subequations}
\begin{align}
	h^{\mu\nu}_{\mathrm{ext}} &\equiv \mathcal{M}(h^{\mu\nu}) = G h_1^{\mu\nu} + G^2 h_2^{\mu\nu} + \calO\left( G^3\right)\,,\label{h1h2}\\
	\psi_{\mathrm{ext}} &\equiv \mathcal{M}(\psi) = G \psi_1 + G^2 \psi_2 + \calO\left(G^3\right)\,.
\end{align}
\end{subequations}
The multipolar expansion is entirely specified by the most general expressions for the multipolar decomposition of the linear coefficients $h_1^{\mu\nu}$ and $\psi_1$. Starting with the tensor coefficient $h_1^{\mu\nu}$ which satisfies the vacuum equations, $\Box h_1^{\mu\nu} = \partial_\nu h_1^{\mu\nu} = 0$, we adopt the same definition of multipole moments as in GR, \textit{i.e.} we write the most general solution in harmonic coordinates as
\be\label{h1mult}	h_{1}^{\mu\nu} = k_1^{\mu\nu} + \partial^\mu \xi_1^\nu + \partial^\nu \xi_1^\mu - \eta^{\mu\nu} \partial_\sigma \xi_1^\sigma \,,\ee
where the gauge vector $\xi_1^\mu$ obeys $\Box \xi_1^\mu = 0$, and where $k_1^{\mu\nu}$ takes a ``canonical'' form~\cite{Thorne:1980ru} in terms of two sets of source multipole moments $I_L(u)$ (mass type) and $J_L(u)$ (current type), with $u\equiv t-r/c$. The moments $I_L$ and $J_L$ are symmetric-trace-free (STF) with respect to the $\ell$ indices composing $L$. We pose 
\begin{subequations}\label{k1mult}
\begin{align}
k_1^{00} &= - \frac{4}{c^2}\sum_{\ell= 0}^{+\infty}\frac{(-)^\ell}{\ell!}\partial_L\left[\frac{I_L(u)}{r}\right]\,,\\
k_1^{0i} &=   \frac{4}{c^3}\sum_{\ell= 1}^{+\infty}\frac{(-)^\ell}{\ell!} \left(\partial_{L-1}\left[\frac{1}{r}\overset{(1)}{I_{iL-1}}(u)\right] + \frac{\ell}{\ell+1}  \partial_{aL-1} \left[ \frac{1}{r}\varepsilon_{iab}J_{bL-1}(u) \right]\right)\,,\\
k_1^{ij} &=  -\frac{4}{c^4}\sum_{\ell= 2}^{+\infty}\frac{(-)^\ell}{\ell!} \left(\partial_{L-2}\left[\frac{1}{r}\overset{(2)}{I_{ijL-2}}(u)\right] + \frac{2\ell}{\ell+1}  \partial_{aL-2} \left[ \frac{1}{r}\varepsilon_{ab(i}\overset{(1)}{J}_{j)bL-2}(u) \right]\right) \,.
\end{align}
\end{subequations}
The moments are defined in such a way that they reduce to the familiar expressions of the Newtonian moments at leading order~\cite{Thorne:1980ru}. The lowest monopolar and dipolar moments satisfy the usual conservation laws $\dd I/\dd t =0$, $\dd^2 I_i/\dd t^2 =0$ and $\dd J_i/\dd t =0$, which are consequences of the harmonic gauge condition $\partial_\nu k_1^{\mu\nu} = 0$. In particular, the monopole $I$ is directly related to the ADM mass $M$ of the system, defined in the usual way at spatial infinity (for either the physical or Einstein metric) by $I=M/\phi_0$.\footnote{Indeed, from $\Box\,h^{\mu\nu} = \frac{16\pi G}{c^{4}}\tau^{\mu\nu}$ we infer that $h^{00} = -\frac{4G I}{c^2} + \cdots$ where $I = c^{-2}\int \dd^3\mathbf{x}\,\tau^{00}$. For a system of point masses which are initially in free motion (when $t\to -\infty$), the ADM mass is given by $M=\sum_A m_A$, while from Eq.~\eqref{taumunu} we see that at spatial infinity $\tau^{00}=T^{00}/\phi_0 = c^2 \sum_A m_A\delta_A/\phi_0$, hence we have $I=M/\phi_0$.} The linear gauge vector $\xi^\mu$ in~\eqref{h1mult} similarly admits a decomposition in terms of four moments called the gauge moments, $W_L$, $X_L$, $Y_L$ and $Z_L$ adopting the same definition as in GR, and given by Eqs.~(37) in~\cite{Blanchet:2013haa}. 

To obtain the expressions of the moments $I_L$, $J_L$ (and also $W_L$, $X_L$, $Y_L$ and $Z_L$) as functions of the source, the procedure is essentially the same as in GR. The multipole expansion of $h^{\mu\nu}$ (noted $\mathcal{M}(h^{\mu\nu})$, as in~\cite{Blanchet:2013haa}) is obtained using a matching to the PN expansion in the near zone of the source as~\cite{Blanchet:1998in}
\begin{subequations}
\begin{align}\label{Mhmunu}
\mathcal{M}(h^{\mu\nu}) &= \mathop{\mathrm{FP}}_{B=0}\,
\Box^{-1}_\mathrm{ret} \Bigl[ \tilde{r}^B \mathcal{M}(\Lambda^{\mu\nu})\Bigr] - \frac{4 G}{c^4}
\sum_{\ell=0}^{+\infty}
\frac{(-)^\ell}{\ell!}\,\partial_L\!\left[\frac{\mathcal{F}_L^{\mu\nu}(u)}{r}\right] \,.
\end{align}
The first term is a solution of the wave equation defined with a specific regularization to cope with the divergence of the multipole expansion when $r\to 0$. The source term $\mathcal{M}(\Lambda^{\mu\nu})$ is multiplied by the regularization factor $\tilde{r}^B \equiv (r/r_0)^B$ where $B\in\mathbb{C}$ and $r_0$ is a regularization length scale, and a Hadamard finite part (FP) procedure is applied on the Laurent expansion of the analytic continuation of the retarded integral when $B\to 0$. The second term in~\eqref{Mhmunu} represents a homogeneous retarded solution parametrized by the multipole moment function (STF in its indices $L$)
\begin{align}\label{FLmunu}
\quad\mathcal{F}_L^{\mu\nu}(u) &= \mathop{\mathrm{FP}}_{B=0}\int
\dd^3\mathbf{x}\,\tilde{r}^B\,\hat{x}_L\,\int_{-1}^1 \dd z
\,\delta_\ell(z)\,\overline{\tau}^{\mu\nu}(\mathbf{x},u+ z r/c)\,,
\end{align}
\end{subequations}
where $\delta_\ell (z) \equiv \frac{(2\ell+1)!!}{2^{\ell+1} \ell!}(1-z^2)^\ell$ is such that $\int_{-1}^{1} \dd z\,\delta_\ell(z) = 1$ and $\lim_{\ell\to+\infty}\delta_\ell (z) = \delta(z)$. Following the notation of~\cite{Blanchet:2013haa}, the function $\mathcal{F}_L^{\mu\nu}(u)$ is expressed in full generality in terms of the PN expansion  $\overline{\tau}^{\mu\nu}$ of the pseudo-tensor $\tau^{\mu\nu}$ defined in~\eqref{taumunu}. From the result~\eqref{Mhmunu}, we can identify the linearized piece $G h_1^{\mu\nu}$ in~\eqref{h1h2} as being essentially given by the second term in~\eqref{Mhmunu}. However, the situation is more complicated because one also has to take into account the harmonic gauge condition which is not satisfied separately by the second term in~\eqref{Mhmunu}. In the end, the source moments follow from the irreducible decomposition of the space-time components $\mathcal{F}_L^{\mu\nu}(u)$. Defining $\Sigma \equiv (\bar\tau^{00} + \bar\tau^{ii})/c^2$, $\Sigma_i \equiv \bar\tau^{0i}/c$ and $\Sigma_{ij} \equiv \bar\tau^{ij}$, we get
\begin{subequations}\label{momentsIJ}
	\begin{align}
	I_L(u) &= \mathop{\mathrm{FP}}_{B=0}\int \dd^3 \mathbf{x} \,\tilde{r}^B \int_{-1}^1 \dd z \bigg[ \delta_\ell(z) \hat{x}_L \Sigma - \frac{4(2\ell +1)}{c^2(\ell+1)(2\ell+3)} \delta_{\ell+1}(z) \hat{x}_{iL} \Sigma^{(1)}_i \nn\\
	& \qquad \qquad \qquad \qquad \qquad \qquad \qquad \qquad + \frac{2(2\ell + 1)}{c^4 (\ell+1)(\ell+2)(2\ell+5)} \delta_{\ell+2}(z) \hat{x}_{ijL} \Sigma^{(2)}_{ij}  \bigg] ( \mathbf{x}, u+z r/c) \,,\\
	J_L(u) &= \mathop{\mathrm{FP}}_{B=0} \int \dd^3 \mathbf{x} \,\tilde{r}^B \int_{-1}^1 \dd z \,\varepsilon_{ab\langle i_\ell} \bigg[ \delta_\ell(z) \hat{x}_{L-1\rangle a} \Sigma_b -\frac{2\ell+1}{c^2(\ell+2)(2\ell+3)} \delta_{\ell+1}(z) \hat{x}_{L-1\rangle ac} \Sigma^{(1)}_{bc}\bigg]( \mathbf{x}, u + z r/c)\,.
	\end{align}
\end{subequations}
These expressions are formally identical as those in GR, but of course in ST theory the source terms $\Sigma$, $\Sigma_i$ and $\Sigma_{ij}$  depend on the scalar field through Eqs.~\eqref{exprLambda}. Similarly, the gauge moments' expressions are formally identical to the GR case and are given by Eqs.~(125) in~\cite{Blanchet:2013haa}.

Next, in ST theory the scalar field brings, in addition to $I_L(u)$ and $J_L(u)$, a new set of multipole moments which we call $I_L^s(u)$, also chosen to be STF. Finding their expression is simpler than in GR (we do not need a correction due to the gauge condition) and the linear piece $G \psi_1$ can be directly identified from the multipole expansion of $\psi$. We find
\begin{subequations}
\begin{align} 
\mathcal{M}(\psi) &= \mathop{\mathrm{FP}}_{B=0}\,
\Box^{-1}_\mathrm{ret} \Bigl[ \tilde{r}^B \mathcal{M}(\Lambda^s)\Bigr] + G \psi_1 \,,\label{Mpsi}\\
\text{with}\quad\psi_1 &= - \frac{2}{c^2}
\sum_{\ell=0}^{+\infty}
\frac{(-)^\ell}{\ell!}\,\partial_L\!\left[\frac{I_L^s(u)}{r}\right]\,.\label{psi1}
\end{align}
\end{subequations}
Similarly to the GR case, the scalar moments are obtained in closed form and we have, defining $\Sigma^s \equiv - \bar\tau_s/c^2$,
\be\label{psi1mult} 
I^s_L(u) = \mathop{\mathrm{FP}}_{B=0} \int \dd^3 \mathbf{x}\,\tilde{r}^B  \int_{-1}^1 \dd z \,\delta_\ell(z) \,\hat{x}_L \Sigma^s ( \mathbf{x}, u+z r/c)\,.
\ee
Note that in contrast to the tensor monopole $I$, the scalar monopole $I^s$ is not constant but its time-variation will be a post-Newtonian effect, \textit{i.e.} $\dd I^s/\dd t = \mathcal{O}(c^{-2})$. Later, we will define $I^s(u) = \phi_0^{-1}[ m^s+\frac{1}{c^2}E^s(u) ]$ where $m^s$ is constant and $E^s(u)$ is the time-varying PN correction, see Eq.~\eqref{Is}.

\subsection{Memory and tail effects in ST theory}

Once the vacuum linearized solutions~\eqref{k1mult} and~\eqref{psi1} are obtained, with explicit expressions for the multipole moments as integrals over the PN expansions of $\tau^{\mu\nu}$ and $\tau_s$, the non-linear contributions can be computed by adapting to ST theories the Multipolar-post-Minkowskian (MPM) algorithm of~\cite{Blanchet:1985sp}. In the following, we will focus our discussion on the new effects specific to ST theories but at the end, we will give the complete non-linear contributions needed to control the waveform to 1.5PN order beyond the quadrupole radiation of GR. 

In GR, the non-linear memory~\cite{Christodoulou:1991cr,Thorne:1992sdb,Blanchet:1992br,Wiseman:1991ss} is a non-local effect due to the radiation of linear waves by the stress-energy tensor, dominantly associated with the mass quadrupole moment. In ST theory, there is a new type of memory effect associated with the scalar dipole radiation that comes from the quadratic interaction between two scalar dipole moments, say $I_i^s \times I_j^s$. As we shall see, such an effect arises at 1.5PN order in the waveform. For the dipole radiation, the linear approximation to the scalar field given by~\eqref{psi1} reads (with $c=G=1$) 
\be\label{psi1dipole}
\psi_1{\Big|}_{I_i^s} = -2 \partial_i\left[\frac{I_i^s(u)}{r}\right] = 2 n^i\Bigl[ r^{-1} \overset{(1)}{I^s_i}(u) + r^{-2} I^s_i(u) \Bigr]\,.\ee
The equation to be solved for the quadratic interaction $I_i^s \times I_j^s$, including the non-linear memory effect, is\footnote{The memory effect arises only in the tensor part $h^{\mu\nu}_2$ and not in the scalar part $\psi_2$, though there will be a local (non-memory) contribution in the scalar field due to the interaction $I_i^s \times I_j^s$, see Eq.~\eqref{radUs} below.} 
\begin{subequations}\label{eqordre2}
\begin{align}
& \Box h^{\mu\nu}_2{\Big|}_{I_i^s \times I_j^s} = \Lambda^{\mu\nu}_2{\Big|}_{I_i^s \times I_j^s}\,,\\
\label{source2}
\text{with }\quad &\Lambda^{\mu\nu}_2{\Big|}_{I_i^s \times I_j^s} \equiv \lambd\left(\eta^{\mu\rho}\eta^{\nu\sigma}-\frac{1}{2}\eta^{\mu\nu}\eta^{\rho\sigma}\right)\partial_\rho\psi_1\partial_\sigma\psi_1{\Big|}_{I_i^s \times I_j^s}\,,
\end{align}
\end{subequations}
where the source term has been obtained from Eq.~\eqref{Lambdas}. Following the MPM algorithm, we shall obtain the solution of this equation, together with the harmonic coordinate condition, by the following two steps process:
\begin{subequations}\label{MPM}
	\begin{align}
	h^{\mu\nu}_2{\Big|}_{I_i^s \times I_j^s} &= u^{\mu\nu}_2 + v^{\mu\nu}_2{\Big|}_{I_i^s \times I_j^s}\,.
	\end{align}
The first step consists in constructing a particular solution $u^{\mu\nu}_2$ of Eq.~\eqref{eqordre2} using the regularized retarded Green's function. However, as it does not satisfy the harmonicity condition it is not an acceptable solution yet. To overcome this difficulty, in the second step, we compute the divergence $w_2^\mu = \partial_\nu u_2^{\mu\nu}$ and then apply the MPM algorithm, symbolized below by the operator $\mathcal{H}$ acting on the vector $w_2^\mu$. More precisely, it consists in constructing a homogeneous solution $v_2^{\mu\nu}$ of the wave equation such that its divergence is exactly the opposite of the divergence of $u_2^{\mu\nu}$, \textit{i.e.} $\Box v_2^{\mu\nu}=0$ and $\partial_\nu v_2^{\mu\nu} = - w_2^{\mu}$. The algorithm  $\mathcal{H}$ is defined by Eqs.~(2.12) of~\cite{Blanchet:1997ji}. To summarize, the two steps read
\begin{align}
\text{(i)}\qquad u^{\mu\nu}_2{\Big|}_{I_i^s \times I_j^s} &= \mathop{\mathrm{FP}}_{B=0}\,
\Box^{-1}_\mathrm{ret} \Bigl[ \tilde{r}^B \Lambda^{\mu\nu}_2\Bigr]{\Big|}_{I_i^s \times I_j^s}\,,\label{u2}\\
\text{(ii)}\qquad v^{\mu\nu}_2{\Big|}_{I_i^s \times I_j^s} &= \mathcal{H}\Bigl[w^{\mu}_2\equiv \partial_\nu u^{\mu\nu}_2\Bigr]{\Big|}_{I_i^s \times I_j^s}\,.
	\end{align}
A useful point is that the divergence of the particular solution $u_2^{\mu\nu}$ comes only from the differentiation of the regularization factor $\tilde{r}^B$ (since the source term obeys $\partial_\nu \Lambda_2^{\mu\nu} =0$), hence
\be\label{w2}
w^{\mu}_2{\Big|}_{I_i^s \times I_j^s} = \mathop{\mathrm{FP}}_{B=0}\,
\Box^{-1}_\mathrm{ret} \Bigl[ B\, \tilde{r}^B \frac{n^i}{r}\Lambda^{\mu i}_2\Bigr]{\Big|}_{I_i^s \times I_j^s}\,.
\ee
\end{subequations}
Due to the presence of the explicit factor $B$, we only have to look at the pole $\propto 1/B$ coming from the integration. It turns out that, even without having controlled the full $u_2^{\mu\nu}$ before, the result~\eqref{w2} is very simple to compute. Inserting the source term~\eqref{source2} computed with~\eqref{psi1dipole}, and integrating by means of Eq.~(A18) of~\cite{Blanchet:1997ji}, we get, to the leading $1/r$ order when $r\to+\infty$,
\begin{align}
w_2^0{\Big|}_{I_i^s \times I_j^s} &= \frac{\lambdprime}{r}\left[
-\frac{4}{3} \overset{(2)}{I}{}^s_i \overset{(2)}{I}{}^s_i + \frac{\dd}{\dd u}\left(\frac{4}{3}\overset{(1)}{I^s_i} \overset{(2)}{I^s_i} - \frac{8}{9}\overset{}{I}{}^s_i \overset{(3)}{I^s_i}\right)\right] + \mathcal{O}\left(\frac{1}{r^2}\right)\,,\\
w_2^i{\Big|}_{I_i^s \times I_j^s} &= \frac{\lambdprime}{r}\frac{\dd}{\dd u}\left(\frac{4}{9} n^i \overset{}{I}{}^s_j \overset{(3)}{I^s_j} + \frac{8}{9} n^j I^s_j \overset{(3)}{I^s_i} - \frac{4}{9} n^j \overset{}{I^s_i}{} \overset{(3)}{I^s_j}\right) + \mathcal{O}\left(\frac{1}{r^2}\right)\,.
\end{align}
Applying the algorithm $\mathcal{H}$ or, rather, its version at leading order $1/r$ explained in Appendix B of~\cite{Blanchet:1997jj}, and keeping only the non-local (hereditary) terms (we shall add all instantaneous terms in the end results below), we get
\be\label{v200}
v_2^{00}{\Big|}_{I_i^s \times I_j^s} = \frac{\lambdprime}{r}\left[\frac{4}{3} \int_{-\infty}^u \dd v \, \overset{(2)}{I^s_i}(v) \overset{(2)}{I^s_i}(v) + \text{``inst.''}\right] + \mathcal{O}\left(\frac{1}{r^2}\right)\,,
\ee
while $v_2^{0i}$ and $v_2^{ij}$ are purely instantaneous. The term~\eqref{v200} has the form of a mass correction, and taking into account $k_1^{00} = -4 I/r+\cdots$ together with the link $I = M/\phi_0$, where $M$ is the constant ADM mass, we obtain the Bondi-type mass taking into account radiation loss as
\be\label{MB}
M_\text{B} (u) = M - \frac{\left(3+2\omega_0 \right)\phi_0}{3} \int_{-\infty}^u \dd v \, \overset{(2)}{I^s_i}(v) \overset{(2)}{I^s_i}(v)\,,
\ee
which implies the standard dipolar mass law in ST theory, as given for instance by Eq.~(9) in~\cite{Bernard:2019yfz}.
 
Let us now consider the piece $u_2^{\mu\nu}$ defined by~\eqref{u2}, and look for non-local effects therein. For any wave equation whose source term is made of quadratic products of linear waves, like Eq.~\eqref{psi1dipole}, we know that the non-local terms arise only from the coefficient $1/r^2$ in the source, see the proof in~\cite{Blanchet:1997ji}. For the dipolar waves, this takes the form of the stress-energy tensor of massless radiation, namely
\begin{align}\label{1surr2}
\Lambda^{\mu\nu}_2 &= \frac{k^\mu k^\nu}{r^2} \sigma(u,\bm{n}) + \mathcal{O}\left(\frac{1}{r^3}\right)\,,
\end{align}
where $k^\mu=(1,n^i)$ is the Minkowski null vector and the energy density in the dipolar wave is given by
\begin{align}\label{sigma}
\sigma(u,\bm{n}){\Big|}_{I_i^s \times I_j^s} &= 4 \lambd n^i n^j \overset{(2)}{I^s_i}(u) \overset{(2)}{I^s_j}(u) \,.
\end{align}
Thus, the energy distribution in the dipolar waves is
\begin{align}\label{dEwavedOmega}
\frac{\dd E^\text{wave}}{\dd \Omega} &= \frac{\phi_0}{16\pi}\int_{-\infty}^u \dd v \,\sigma(v,\bm{n})\,,
\end{align}
with the total energy in the waves given by $\Delta E^\text{wave} \equiv \int\dd\Omega\,\frac{\dd E^\text{wave}}{\dd \Omega} = M - M_\text{B}$. As we  have already mentionned, the only hereditary contributions in $u_2^{\mu\nu}$ comes from the piece~\eqref{1surr2}. At this stage, it is convenient to perform a gauge transformation with gauge vector~\cite{Blanchet:1986dk}
\begin{align}\label{zeta2}
\zeta^{\mu}_2 \equiv \mathop{\mathrm{FP}}_{B=0}\,
\Box^{-1}_\mathrm{ret} \Bigl[ \tilde{r}^B \frac{k^\mu}{2 r^2} \int_{-\infty}^u \dd v \,\sigma(v,\bm{n})\Bigr]\,.
\end{align}
Posing ${u'}^{\mu\nu}_2=u^{\mu\nu}_2+\partial^\mu\zeta^{\nu}_2+\partial^\nu\zeta^{\mu}_2-\eta^{\mu\nu}\partial_\rho\zeta^{\rho}_2$ and adding Eq.~\eqref{v200}, we then obtain the metric in the new gauge, \textit{i.e.} ${h'}^{\mu\nu}_2 \equiv {u'}^{\mu\nu}_2 + {v}^{\mu\nu}_2$, in the form
\begin{align}\label{h'munu2}
{h'}^{\mu\nu}_2 = \frac{1}{r}\left[\int_{-\infty}^u \dd v \,\mathcal{H}^{\mu\nu}_2(v,\bm{n}) + \text{``inst.''} \right] + \mathcal{O}\left(\frac{1}{r^2}\right) \,,
\end{align}
where the non-local contributions read
\begin{subequations}\label{Hmunu2}
\begin{align}
\mathcal{H}^{00}_2{\Big|}_{I_i^s \times I_j^s} &= \frac{2 \lambd}{3} \,\overset{(2)}{I^s_i}\overset{(2)}{I^s_i}\,,\\
\mathcal{H}^{0i}_2{\Big|}_{I_i^s \times I_j^s} &= \frac{2 \lambd}{3} \left[n^{ijk}\overset{(2)}{I^s_j}\overset{(2)}{I^s_k} + n^{i}\overset{(2)}{I^s_j}\overset{(2)}{I^s_j} - n^{j}\overset{(2)}{I^s_i}\overset{(2)}{I^s_j} \right]\,,\\
\mathcal{H}^{ij}_2{\Big|}_{I_i^s \times I_j^s} &= \lambd \left[ n^{ijkl}\overset{(2)}{I^s_k}\overset{(2)}{I^s_l} + \frac{1}{3} n^{ij} \overset{(2)}{I^s_k}\overset{(2)}{I^s_k} - \frac{1}{3} \delta^{ij} n^{kl} \overset{(2)}{I^s_k}\overset{(2)}{I^s_l} - \frac{2}{3} \overset{(2)}{I^s_i}\overset{(2)}{I^s_j} + \frac{1}{3} \delta^{ij}\overset{(2)}{I^s_k}\overset{(2)}{I^s_k} \right]\,.
\end{align}
\end{subequations}
It is then straightforward to see that this implies a correction in the observable radiative quadrupole moment (defined  in Eq.~\eqref{GWFs} below) due to the dipolar memory effect as given by
\begin{align}\label{deltaUmemory}
\delta \mathcal{U}^\text{memory}_{ij}{\Big|}_{I_i^s \times I_j^s} = \frac{\lambdprime}{3} \int_{-\infty}^u \dd v \, \overset{(2)}{I^s_{\langle i}}(v) \overset{(2)}{I^s_{j\rangle}}(v) \,.
\end{align}


Focusing now on the dipolar tail effect, it arises from the quadratic interaction between the conserved gravitational monopole $I=M/\phi_0$ and the scalar dipole moment $I_i^s(u)$. At second order, we have to solve, for this particular interaction $I \times I_j^s$, using Eqs.~\eqref{lambdas} and~\eqref{psi1dipole},
\begin{align}\label{Boxpsi2}
\Box \psi_2{\Big|}_{I \times I_j^s} =  \frac{8 I}{r^2} \,n^i \overset{(3)}{I^s_i}(u) + \mathcal{O}\left(\frac{1}{r^3}\right)\,.
\end{align}
Using Eq.~(A8b) in~\cite{Blanchet:1997ji}, we obtain
\begin{align}\label{psi2tail}
\psi_2^\text{tail}{\Big|}_{I \times I_j^s} =  \frac{4 I}{r} \left[ n^i \int_{-\infty}^u \dd v \,\overset{(3)}{I^s_i}(v) \ln\left(\frac{u-v}{2 r}\right)+ \text{``inst.''} \right] + \mathcal{O}\left(\frac{1}{r^2}\right)\,.
\end{align}

Finally, we present the complete results for the observable radiative moments including memory and tail contributions up to the 1.5PN order, as well as all instantaneous terms. We introduce a radiative type coordinate system $(T, R)$, with $U \equiv T - R/c$ being an asymptotically null coordinate such that
\begin{equation}\label{Uu}
U = u - \frac{2 G I}{c^3}\ln\left(\frac{r}{c b}\right) + \mathcal{O}\left(\frac{1}{r}\right)\,,
\end{equation}
where $I$ is the mass monopole moment and $b$ is an arbitrary constant time-scale. Then, we denote $\mathcal{U}_L$, $\mathcal{V}_L$ and $\mathcal{U}^s_L$ the radiative moments and parametrize the asymptotic transverse-tracefree (TT) tensorial waveform and the scalar waveform in the radiative coordinate system at leading order $1/R$ in the distance. We have\footnote{We restore the factors $G$ and $c$. In~\eqref{GWFa} we denote by $h_{ij}^\text{TT}$ the TT projection of the gothic spatial metric deviation, which differs by a sign from the ordinary (covariant) spatial metric deviation.}
\begin{subequations}\label{GWFs}
\begin{align}
 	h_{ij}^\text{TT} &= - \frac{4G}{c^2 R} \perp^\text{TT}_{ijab} \sum_{\ell=2}^{+\infty} \frac{1}{c^\ell \ell!}\Big( N_{L-2}\,\mathcal{U}_{ab L-2}(U) - \frac{2\ell}{c(\ell+1)} N_{c L-2} \varepsilon_{cd(a}\mathcal{V}_{b)dL-2}(U)\Big) + \mathcal{O}\Big(\frac{1}{R^2}\Big)\,,\label{GWFa}\\
 	\psi &= - \frac{2G}{c^2 R}\sum_{\ell=0}^{+\infty} \frac{1}{c^\ell \ell!} N_L \mathcal{U}_L^s(U) + \mathcal{O}\Big(\frac{1}{R^2}\Big)\,, 
\end{align}
\end{subequations}
where $\perp^\text{TT}_{ijab}\equiv\frac{1}{2}(\perp_{ia}\perp_{jb}+\perp_{ja}\perp_{ib}-\perp_{ij}\perp_{ab})$ with $\perp_{ij}=\delta_{ij}-N_iN_j$ being the TT projection operator (we remind that \textit{e.g.} $N_{L-2}=N_{i_1}\cdots N_{i_{\ell-2}}$). Since the matter fields are minimally coupled to the physical metric $g_{\mu\nu}=\varphi^{-1}\tilde{g}_{\mu\nu}$, the GW detector will respond to the tidal field associated with the physical metric, \textit{i.e.} the linearized Riemann tensor of the physical metric. Thus, the separation vector between the entry and end mirrors of one arm of a laser-interferometric detector will obey the standard GR law at linear order, $\ddot{\xi} = c^{2} R^\text{lin}_{0i0j}\,\xi^j$, but with the Riemann tensor being in ST theory related to both the tensorial and scalar waveforms~\eqref{GWFs}. We have, see Eq.~(18) in~\cite{Sennett:2016klh},
\be
c^{2}R^\text{lin}_{0i0j} = \frac{1}{2} \ddot{h}_{ij}^\text{TT} + \frac{1}{2} \!\perp_{ij}\!\ddot{\psi}\,,
\ee
showing explicitly the decomposition of the detector's response into tensorial and scalar modes. 

We now  present the results for the radiative type moments. The tensorial and scalar fluxes $\mathcal{F}$ and $\mathcal{F}^s$ are deduced in terms of them directly from the waveforms~\eqref{GWFs} as
\begin{subequations}\label{fluxes}
\begin{align}
\mathcal{F} &= \frac{c^3 R^2 \phi_0}{32 \pi G} \int \!\dd\Omega \left(\frac{\partial h_{ij}^\text{TT}}{\partial U}\right)^2 \!\!= \sum_{\ell=2}^{+\infty} \frac{G \phi_0}{c^{2\ell+1}}\Bigg( \frac{(\ell+1)(\ell+2)}{(\ell-1) \ell \ell! (2\ell+1)!! }  \overset{(1)}{\mathcal{U}}_L\overset{(1)}{\mathcal{U}}_L + \frac{4 \ell (\ell+2)}{c^2 (\ell-1) (\ell+1)! (2\ell+1)!!} \overset{(1)}{\mathcal{V}}_L\overset{(1)}{\mathcal{V}}_L\Bigg) \,,\\
\mathcal{F}^s &= \frac{c^3 R^2 \phi_0 \lambd}{16\pi G} \int\!\dd\Omega \left( \frac{\partial \Psi}{\partial U}\right)^2 = \sum_{\ell=0}^{+\infty} \frac{G \phi_0 \lambd}{c^{2\ell+1} \ell! (2\ell+1)!!}\overset{(1)}{\mathcal{U}^s}_L\overset{(1)}{\mathcal{U}^s}_L\,.
\end{align}
\end{subequations}
To 1.5PN order, only the mass quadrupole radiative moment differs from its twice-differentiated source counterpart,
\begin{align}\label{radUij}
\mathcal{U}_{ij} = \overset{(2)}{I_{ij}} &+ \frac{2G M}{\phi_0 c^3} \int_{-\infty}^U \!\dd V \overset{(4)}{I_{ij}}(V) \left[ \ln{\left(\frac{U-V}{2b} \right)} + \frac{11}{12} \right]\nn \\
&+ \frac{G\lambd}{3c^3}\Bigg( \int_{-\infty}^U \!\dd V \overset{(2)}{I^s_{\langle i}}(V)\overset{(2)}{I^s_{j\rangle}}(V) -I^s_{\langle i}\overset{(3)}{I^s_{j\rangle}}- \overset{(1)}{I^s_{\langle i}}\overset{(2)}{I^s_{j\rangle}} -\frac{1}{2}I^s \overset{(3)}{I^s_{ij}} \Bigg)+ \bigO5 \,.
 \end{align}
The difference from $\overset{(2)}{I_{ij}}$ is made of two contributions, both of them being at 1.5PN order: 
\begin{itemize}
\item[(i)] the usual mass quadrupole tail correction in GR with the usual instantaneous term with coefficient $\frac{11}{12}$, except that the mass monopole $I$ therein is related to the ADM mass by $I=M/\phi_0$;
\item[(ii)] the dipolar memory effect found in Eq.~\eqref{deltaUmemory}, to which we have added the corresponding instantaneous contributions obtained by a detailed calculation.
\end{itemize}
In addition, we find that for the 1.5PN ST waveform the radiative type scalar monopole, dipole and quadrupole moments acquire some important tail contributions, namely
\begin{subequations}\label{radUUiUijs}
\begin{align}
\mathcal{U}^s &= I^s + \frac{2 G M}{\phi_0 ^2 c^5} \int_{-\infty}^U \dd V \overset{(2)}{E^s}(V)\ln\left(\frac{U-V}{2b}\right)  + \frac{G}{ c^5}\left(1-\frac{\phi_0 \omega_0'}{3+2\omega_0}\right) \Bigg[\frac{2}{9} I_k^s\overset{(3)}{I_k^s}-2 I^s \frac{\overset{(1)}{E^s}}{\phi_0}\Bigg]+\bigO7 \,,\label{radUs}\\
\mathcal{U}_i^s &= \overset{(1)}{I_i^s} + \frac{2GM}{\phi_0 c^3} \int_{-\infty}^U \dd V \overset{(3)}{I_i^s}(V) \left[ \ln{\left(\frac{U-V}{2b} \right)} + 1 \right] \nn\\
&\qquad\ +\frac{G}{c^5} \Bigg[ -\frac{1}{5}\overset{(1)}{I^s_{k}}\overset{(3)}{I_{ik}} -\frac{1}{5}\overset{(2)}{I^s_{k}}\overset{(2)}{I_{ik}} +\frac{3}{5}\overset{(3)}{I^s_{k}}\overset{(1)}{I_{ik}}  +\frac{3}{5}\overset{(4)}{I^s_{k}}I_{ik} 
- \varepsilon_{iab}J_a \overset{(3)}{I^s_b}
- 4\overset{(1)}{W}\overset{(2)}{I^s_i}
- 4 \overset{(2)}{W}\overset{(1)}{I^s_i}
+ 4 I^s \overset{(2)}{Y_i} \nn\\[-0.1cm]
&\qquad\qquad\quad\ + \Big( 1-\frac{\phi_0\omega_0'}{3+2\omega_0}\Big)\Bigg( 
-2 \frac{\overset{(1)}{E_s} }{\phi_0}\overset{(1)}{I^s_i} 
-2 \frac{\overset{(2)}{E_s} }{\phi_0}I^s_i 
+ \frac{2}{5} \overset{(3)}{I^s_k}\overset{(1)}{I^s_{ik}} 
+ \frac{2}{5} \overset{(4)}{I^s_k}I^s_{ik} \Bigg)\Bigg] + \bigO7\,, \\
\mathcal{U}_{ij}^s &= \overset{(2)}{I_{ij}^s}(U)  + \frac{2 G M}{\phi_0 c^3} \int_{-\infty}^U \dd V \overset{(4)}{I^s_{ij}}(V)\left[ \ln{\left(\frac{U-V}{2b} \right)} + \frac{3}{2} \right]  - \frac{G}{c^3} I^s \overset{(3)}{I_{ij}}  +\bigO5 \,.
\end{align}
\end{subequations}
In these expressions we have conveniently posed, recalling the definition of the sensitivity parameters~\eqref{sAk}, 
\begin{subequations}\label{Is}
\begin{align}
& I^s(u) = \frac{1}{\phi_0}\biggl[ m^s+\frac{E^s(u)}{c^2} \biggr]\,,\\
\text{with}\qquad & m^s = -\frac{1}{\lambdprime}\sum_A m_A\bigl(1-2s_A\bigr)\,.
\end{align}
\end{subequations}
The Newtonian value $m^s$ in the limit $c\to+\infty$ is constant and represents the total ``scalar charge'' of the system; $m^s$ is zero for binary black holes.

We can already make a few observations on Eqs.~\eqref{radUUiUijs}. First, these expressions are valid in the frame of the centre-of-mass (CM) of the system defined by $I_i=0$. Then, we notice the coupling between the tensorial moments and the scalar ones through some instantaneous terms in $\mathcal{U}_{i}^s$ and $\mathcal{U}_{ij}^s$. Finally, in $\mathcal{U}_{i}^s$, the gauge moments $W$ and $Y_i$ coming from the tensorial sector, see Eq.~\eqref{h1mult}, appear and are coupled to the scalar mass monopole and dipole. In GR the gauge moments do not contribute to the waveform before the 2.5PN order while in ST theory they already contribute to the flux at 1.5PN order and to the scalar waveform at 2PN order. Besides Eqs.~\eqref{radUUiUijs}, all other radiative moments are linearly related to the source moments \textit{via} the usual formul{\ae}, \textit{e.g.} $\mathcal{U}_L^s = \dd^\ell I_L^s/\dd u^\ell$.  
 
\section{The scalar-tensor multipole moments and flux of compact binaries}
\label{sec:compactbin}


Now that the general formalism has been laid out, we apply it to the case of a non-spinning compact binary system to compute the \textit{source} multipole moments, and subsequently, the \textit{radiative} ones. Consistently with the general formul{\ae} presented in the previous section, we focus on the ST fluxes at 1.5PN order beyond the GR's quadrupole formula (\textit{i.e.} 2.5PN beyond the leading dipolar order of ST theories). For two point-particles of masses $m_A(\phi)$, the stress-energy tensor deriving from Eq.~\eqref{matteract} reads
\be
T^{\mu \nu}(t,\mathbf{x}) = \sum _{A=1,2} \frac{m_A(\phi) \,v_A^\mu v_A^\nu}{\sqrt{-(g_{\alpha \beta})_A \frac{v_A^\alpha v_A^\beta}{c^2} }}\frac{\delta_A}{\sqrt{-g}}\,.
\ee
Our notation for point-particle systems is given in Sec.~\ref{sec:not}. From the explicit dependence of the stress-energy tensor on the scalar field through the masses we obtain
\be\label{dTdphi}
T-2\varphi\frac{\partial T}{\partial \varphi} = - c \sum _{A=1,2} m_A\bigl(1-2s_A\bigr) \sqrt{-(g_{\alpha \beta})_A \,v_A^\alpha v_A^\beta } \,\frac{\delta_A}{\sqrt{-g}}\,.
\ee
Next, we define the following compact support source densities\footnote{Note a change of definition for $\sigma_s$: in Ref.~\cite{Bernard:2018hta}, the definition given in Eq.~(4.16) should not be used. Indeed, this definition was never used during the computation in~\cite{Bernard:2018hta}; instead, only the 1PN expansion given in Eq.~(B4) was used, and is correct. Our definition~\eqref{sigmas} for $\sigma_s$ yields~(B4) in~\cite{Bernard:2018hta} at the 1PN approximation.}
\begin{subequations}
	\begin{align}
&	\sigma = \frac{1}{\phi_0 \varphi^3}\frac{T^{00} + T^{ii}}{c^2}\,,\qquad	\sigma_i = \frac{1}{\phi_0 \varphi^3}\frac{T^{0i}}{c} \,,\qquad	\sigma_{ij} = \frac{1}{\phi_0 \varphi^3}T^{ij} \,,\\[5pt]
&	\sigma_s = - \frac{1}{c^2\phi_0}\frac{\sqrt{-g}}{\sqrt{\lambd(3+2\omega)}} \left( T - 2\varphi \frac{\partial T}{\partial \varphi} \right) \,.\label{sigmas}	\end{align}
\end{subequations}
At 1.5PN order, we parametrize the metric and scalar perturbation fields as
\begin{subequations}
\begin{align}
h^{00} &= - \frac{4}{c^2}V - \frac{2}{c^4}\Big(\W + 4 V^2 \Big) + \bigO6 \,,\\
h^{0i} &= - \frac{4}{c^3} V_i  + \bigO{5} \,,\\
h^{ij} &= -\frac{4}{c^4}\Big( \W_{ij} - \frac{1}{2}\delta_{ij}\W \Big) + \bigO{6} \,,\\
\psi &= - \frac{2}{c^2} \psip{0} + \frac{2}{c^4} \Big( 1 - \frac{\omega'_0 \phi_0}{3+2\omega_0}  \Big) \psip{0}^2  +  \bigO6 \,,
\end{align}
\end{subequations}
where we have introduced (extending the usual practice in GR) the elementary potentials obeying the following wave equations
\begin{subequations}
	\begin{align}
	\Box V &= -4\pi G \,\sigma\,,\\
	\Box V_i &= -4\pi G \,\sigma_i\,,\\
	\Box \W_{ij} &= -4\pi G \,\bigl( \sigma_{ij} - \delta_{ij} \sigma_{kk}\bigr) - \partial_i V \partial_j V - \lambd \partial_i \psip{0} \partial_j \psip{0}\,,\\
	\Box \psip{0} &= 4\pi G \,\sigma_s\,.
	\end{align}
\end{subequations}
Finally, inserting these decompositions into the tensor and scalar multipole moments as given by Eqs.~\eqref{momentsIJ} and~\eqref{psi1mult}, we can compute the multipole moments using a series of known techniques; see~\cite{MHLMFB20} for a recent review. Fortunately, at the 1.5PN order, this is relatively easy as we do not need to worry about the subtleties associated with the different regularizations schemes (be it ultraviolet or infrared). The main difficulty is the long calculation of the scalar monopole and dipole moments $I^s$ and $I_i^s$ which are required at 2.5PN order.


With the ST parameters defined at the beginning of the paper (Sec.~\ref{sec:not}), we obtain the source ST multipole moments for compact binaries that are relevant for the final 1.5PN waveform and flux. For example, the tensor quadrupole moment $I_{ij}$ is accurate to order 1.5PN and we find, after reduction to the CM frame,\footnote{For convenience, when working in the CM frame, we denote $r\equiv r_{12}$ and $n^i\equiv n_{12}^i$ the distance and separation vector between the two bodies ($r$ should not be confused with the harmonic-coordinate distance to the source used in Sec.~\ref{sec:MPNinST}).}
\begin{align}\label{Iij}
I_{ij} &= \frac{m \nu r^2}{\phi_0} \Bigg[n^{\langle i} n^{j\rangle}+ \frac{1}{42 c^2} \Bigg\{n^{\langle i} n^{j\rangle} \left(\frac{\tilde{G}\alpha m}{r}\right) (-30 + 48 \nu)  \\
&\qquad\qquad\qquad\qquad\qquad\quad+ n^{\langle  i} n^{j\rangle}v^2 (29 - 87 \nu)  +  n^{\langle i}  v^{j\rangle} (nv) (-24 + 72 \nu)  +  v^{\langle i} v^{j\rangle} (22 - 66 \nu)  \Bigg\}\Bigg] + \bigO4\,.\nn
\end{align}
The other relevant tensor moments (including the two gauge moments $W$ and $Y_i$) are given in Eqs.~\eqref{tensormoments} of Appendix~\ref{app:moments}.

As for the scalar moments, the full expressions at the highest necessary PN order are also given in Appendix~\ref{app:moments}. Hereafter, we only give the source moments at a high enough PN order to compute the hereditary terms. Indeed, the radiative moment $\mathcal{U}_{ij}$ given by Eq.~\eqref{radUij} corrects the source moment by the relevant 1.5PN quadrupole tail effect which is the same as in GR, and by the non-linear memory effect associated with the dipole moment. We will include these effects, applied to binaries, into the final waveform. The necessary source moments to compute hereditary terms are then
\begin{subequations}
\begin{align}
I^s_i =& \ \frac{\alpha^{1/2} \zeta m \nu r }{(1 -  \zeta) \phi_0} \Bigg[-2  \mathcal{S}_{-}  n^{i}  + \frac{n^{i}}{5 c^2} \Bigg\{ \left(\frac{\tilde{G}\alpha m}{r}\right) \Big(9  \mathcal{S}_{-} - 20 \bar{\beta}_{+}  \bar{\gamma}^{-1} \mathcal{S}_{-} - 20 \bar{\beta}_{-}  \bar{\gamma}^{-1} \mathcal{S}_{+} - 13  \mathcal{S}_{-} \nu \nn\\ 
&\qquad\qquad\qquad\quad + \delta \Big[ 20 \bar{\beta}_{-}  \bar{\gamma}^{-1} \mathcal{S}_{-} - 4  \mathcal{S}_{+} + 20 \bar{\beta}_{+}  \bar{\gamma}^{-1} \mathcal{S}_{+}\Big]\Big)
+  v^2  \bigl(-  \mathcal{S}_{-} + 7  \mathcal{S}_{-} \nu - 4  \mathcal{S}_{+} \delta  \bigr) \Bigg\} \nn\\
&\qquad\qquad\qquad\quad 
+ \frac{v^{i}  (nv) }{5 c^2} \Bigg\{ -2  \mathcal{S}_{-}  + 4  \mathcal{S}_{-} \nu + 2  \mathcal{S}_{+} \delta\Bigg\} \Bigg] + \bigO3\,, \\
E^s =& \frac{\alpha^{1/2} \zeta m \nu}{6 (1 -  \zeta)}  \Bigg[\left(\frac{\tilde{G}\alpha m}{r}\right) \Big(-48 \bar{\beta}_{-}  \bar{\gamma}^{-1} \mathcal{S}_{-} + 14  \mathcal{S}_{+} - 48 \bar{\beta}_{+}  \bar{\gamma}^{-1} \mathcal{S}_{+} - 2  \mathcal{S}_{-} \delta \Big) +v^2\Big( \mathcal{S}_{+} -   \mathcal{S}_{-} \delta \Big)   \Bigg] \  + \mathcal{O}\left(\frac{1}{c}\right)\,, \\
I^s_{ij} =&\  \frac{\alpha^{1/2} \zeta m  r^2 \nu n^{i} n^{j}}{(1 -  \zeta) \phi_0}\Big(- \mathcal{S}_{+} + \mathcal{S}_{-} \delta\Big)   + \mathcal{O}\left(\frac{1}{c}\right) \,,
\end{align}
\end{subequations}
together with the Newtonian limit of~\eqref{Iij} and the Newtonian energy $E = \frac{1}{2} m\nu v^2-\frac{\tilde{G} \alpha m^2\nu}{r} + \mathcal{O}(c^{-1})$.

So far, only the ST equations of motion to 1.5PN order were required. However, in order to compute the ST fluxes, we have to differentiate the moments with respect to time, which then uses the ST equations of motion to 2.5PN order. All the conservative (PN-even) terms of the acceleration are available in Ref.~\cite{Bernard:2018hta}, in which the ST equations of motion were obtained up to 3PN order. Concerning the dissipative (PN-odd) terms at 1.5PN and 2.5PN orders, they are given in Ref.~\cite{Mirshekari:2013vb} as functions of the EW multipole moments in harmonic coordinates~\cite{Epstein:1975}. In Appendix~\ref{app:EOM}, we present for the first time the explicit expressions for these PN-odd terms in the equations of motion to 2.5PN order.

Finally, we have checked that all our results concerning the moments (in this section and in Appendix~\ref{app:moments}) are in agreement with the results of Refs.~\cite{Lang:2013fna,Lang:2014osa}. For the comparison, we have to carefully take into account the link between the STF moments and the Epstein-Wagoner moments~\cite{Epstein:1975} used in~\cite{Lang:2013fna,Lang:2014osa}. The Sec.~\ref{sec:EW} in Appendix~\ref{app:moments} gives the required relations between these moments. 

The scalar and tensorial energy fluxes, respectively denoted $\mathcal{F}^s$ and $\mathcal{F}$, are given in terms of the STF radiative multipole moments $\mathcal{U}_L$, $\mathcal{V}_L$ and $\mathcal{U}^s_L$ according to~\eqref{fluxes}. We further split them into instantaneous and tail contributions, following the contributions of the tail integrals in~\eqref{radUij}--\eqref{radUUiUijs},
\begin{subequations}
\begin{align}
\mathcal{F} &= \mathcal{F}_\text{inst} + \mathcal{F}_\text{tail}\,,\\
\mathcal{F}^s &= \mathcal{F}^s_\text{inst} + \mathcal{F}^s_\text{tail}\,.
\end{align}
\end{subequations}
Note that the dipolar memory contribution, given by the last integral in~\eqref{radUij}, becomes instantaneous in the flux and as a consequence, its contribution is included into the instantaneous part. On the other hand, it is convenient to keep some instantaneous terms, such as the one related to the constant $11/12$, into the definition of the tail terms. 

With this caveat in mind, we define 
\begin{subequations}\label{Ftail}
	\begin{align}
	\mathcal{F}_\text{tail} &= \frac{4G^2 M}{5c^8} \overset{(3)}{I_{ij}} \int_{-\infty}^U \!\dd V \overset{(5)}{I_{ij}}(V) \left[ \ln{\left(\frac{U-V}{2b} \right)} + \frac{11}{12} \right] \,,\\
	\mathcal{F}^s_\text{tail} &= \frac{4G^2 M \lambd}{c^6} \Biggl\{ \frac{1}{3}\overset{(2)}{I^s_{i}} \int_{-\infty}^U \!\dd V \overset{(4)}{I^s_{i}}(V) \left[ \ln{\left(\frac{U-V}{2b} \right)} + 1 \right]  +\frac{\overset{(1)}{E^s}}{c^2 \phi_0^2}
	 \int_{-\infty}^U \!\dd V \overset{(3)}{E^s}(V) \ln{\left(\frac{U-V}{2b} \right)}\nn\\ 
	& \qquad\qquad\qquad\qquad\qquad + \frac{1}{30 c^2}\overset{(3)}{I^s_{ij}} \int_{-\infty}^U \!\dd V \overset{(5)}{I^s_{ij}}(V) \left[ \ln{\left(\frac{U-V}{2b} \right)} + \frac{3}{2} \right]\Biggr\}\,.
	\end{align}
\end{subequations}
The complete expression of the instantaneous tensorial flux, valid for general orbits in the CM frame, reads
\begin{align}\label{tensorflux}\nn
\mathcal{F}_\text{inst} &=  \frac{4 \left(2 + \bar{\gamma}\right)}{15c^5} \left(\frac{\tilde{G}\alpha m}{r}\right)^3 \frac{m \nu^2}{r} \Bigg\{-11 (nv)^2 + 12 v^2\\
&+ \frac{1}{28 c^2} \Bigg[ \Bigl(16 - 64 \nu\Bigr)\left(\frac{\tilde{G}\alpha m}{r}\right)^2 + \Bigl(2061 + 840 \bar{\gamma} - 1860 \nu\Bigr) (nv)^4  \\
& \qquad\qquad +  \biggl (2936 + 1344 \bar{\beta}_{+}+ 1120 \bar{\gamma} - 1344 \bar{\beta}_{-} \delta - 120 \nu\biggl)\left(\frac{\tilde{G}\alpha m}{r}\right) (nv)^2 \nn\\
&\qquad\qquad+ \biggl (-2720 -1344 \bar{\beta}_{+} - 1008 \bar{\gamma} + 1344 \bar{\beta}_{-} \delta + 160 \nu\biggl) \left(\frac{\tilde{G}\alpha m}{r}\right) v^2 \nn\\
&\qquad\qquad+ \Bigl(-2974 - 1232 \bar{\gamma} + 2784 \nu\Bigr) (nv)^2 v^2 + \Bigl(785 + 336 \bar{\gamma} - 852 \nu\Bigr) v^{4}\Bigg]\nn\\
&+\frac{1}{12c^3} \Bigg[ \Big(-2 \bar{\gamma} + 48 \zeta \mathcal{S}_{-}^2 \nu\Big) \left(\frac{\tilde{G}\alpha m}{r}\right)^2 (nv) + 66 \bar{\gamma} \left(\frac{\tilde{G}\alpha m}{r}\right) (nv)^3 - 70 \bar{\gamma} \left(\frac{\tilde{G}\alpha m}{r}\right) (nv) v^2\Bigg] \Bigg\} \,.
\end{align}
The tensorial energy flux $\mathcal{F}_\text{inst}$ is in complete agreement with the result of Refs.~\cite{Lang:2013fna,Lang:2014osa}. 

We also compute the instantaneous scalar flux $\mathcal{F}^s_\text{inst}$ to 1.5PN order, complementing by a half post-Newtonian order the previous result by Lang~\cite{Lang:2014osa}. As its full expression in the center of mass is very long, we have preferred to relegate it to Appendix~\ref{app:fluxcoeffs}.

However, when comparing our scalar flux with the one obtained by Ref.~\cite{Lang:2014osa}, we have found a discrepancy at~1PN order that could not be resolved, despite the fact that we agree on all the ST multipole moments separately, notably the~2PN monopole and dipole scalar moments. In order to investigate this disagreement, we have computed the~1.5PN scalar waveform and have found that it is in perfect agreement with the scalar waveform presented in Eqs.~(5.2) of~Ref.~\cite{Lang:2014osa}. We have then computed the flux: (i) by integrating the scalar waveform following Eq.~(6.6) of Ref.~\cite{Lang:2014osa}; (ii) with the direct formula given by Eq.~\eqref{FscalarEW} of Appendix~\ref{app:moments}, where the EW moments were replaced by their center-of-mass expressions as given by Eqs.~(3.50) of Ref.~\cite{Lang:2014osa} and Eqs.~(5.10) of Ref.~\cite{Lang:2013fna}. In both cases, we recover the scalar flux as given in Appendix~\ref{app:fluxcoeffs}, and not the scalar flux of Ref.~\cite{Lang:2014osa}. The explicit difference between our scalar flux and the scalar flux of  Ref.~\cite{Lang:2014osa} is given explicitly in Eq.~\eqref{DiffLang} of Appendix~\ref{app:fluxcoeffs}, along with the full expression for the scalar flux.
\section{Reduction to quasi-circular orbits}
\label{sec:circular}

For quasi-circular orbits, the expressions of the fluxes simplify considerably and one can work out explicitly the tail terms~\eqref{Ftail} from standard methods. The usual frequency dependent PN variable which permits to obtain gauge invariant results in GR is easily generalized to ST theories as
\be\label{x}
x \equiv \left(\frac{\tilde{G} \alpha m \omega}{c^3}\right)^{2/3}\,, 
\ee
where $\omega = 2\pi/P$ is the orbital frequency of the quasi-circular orbit, with $P$ the period. It is related to the orbital separation $r$ and to the gauge dependent PN variable $\gamma \equiv \frac{\tilde{G} \alpha m}{c^2 r}$, by the sum of Eqs.~(5.2)--(5.3) of~\cite{Bernard:2018ivi} which includes all contributions up to relative 3PN order, including the dipolar tail term at 3PN order. For the present work, we only need this relation to relative 2PN order, which we reproduce here for convenience, %
\begin{align}\label{omegaInst}
\omega^{2} &= \frac{\alpha \tilde{G} m}{r^3} \Biggl(1+ \gamma \bigg\{-3- 2 \bar\beta_{+}+ \nu-  \overline{\gamma}+ 2 \bar\beta_{-} \delta\bigg\} \nn\\
&\qquad\quad\quad
+ \gamma^2 \bigg\{ 6 + 8 \bar\beta_{+} - 2 \bar\chi_{+} + \bar\delta_{+}  + \bigl(5 + 2 \bar\beta_{+}\bigr) \overline{\gamma} + \frac{5}{4} \overline{\gamma}^2 + \delta \Big[2 \bar\chi_{-} + \bar\delta_{-} - 2 \bar\beta_{-} \bigl(4 + \overline{\gamma}\bigr)\Big]  \nonumber\\&\qquad\quad\qquad\ \,\qquad + \nu \Big[\frac{41}{4} + \bar\beta_{+} + 4 \bar\chi_{+} - 2 \bar\delta_{+} + 24 \bar\beta_{-}^2 \overline{\gamma}^{-1}   - 24 \bar\beta_{+}^2 \overline{\gamma}^{-1} + 5 \overline{\gamma}-  \frac{1}{2} \overline{\gamma}^2 \Big]  + 5 \bar\beta_{-} \nu \delta  + \nu^2 \bigg\} 
\Biggr) \,.
\end{align}
The expression of $\gamma$ in terms of $x$ at 2PN order is deduced by inversion of this expression, and is also given to 3PN order in~\cite{Bernard:2018ivi}. Our final results for the tensorial and scalar fluxes at relative 2PN order expressed in terms of the gauge invariant variable~\eqref{x} read
\begin{subequations}\label{circular}
\begin{align}
\mathcal{F}_\text{circ} &= \frac{32 c^5 x^5 \nu^2 (1+\bar{\gamma}/2)}{5\tilde{G} \alpha}\Bigg(1  + \frac{x}{336}\Bigg\{ -1247 - 896 \bar{\beta}_{+} - 448 \bar{\gamma} + 896 \bar{\beta}_{-} \delta - 980 \nu \Bigg\}  + 4\pi \left(1+\frac{\bar{\gamma}}{2}\right) x^{3/2}\Bigg)\,,\\
\mathcal{F}^s_\text{circ} &= \frac{c^5 x^5  \nu^2  \zeta }{3\tilde{G} \alpha} \Bigg(4  \mathcal{S}_{-}^2 x^{-1} \nn \\
& \qquad\qquad\qquad +\frac{1}{15} \Bigg\{-24 \zeta^{-1} \bar{\gamma} - 120 \mathcal{S}_{-}^2 - 80 \bar{\beta}_{+} \mathcal{S}_{-}^2 - 40 \bar{\gamma} \mathcal{S}_{-}^2 + 240 \bar{\beta}_{+} \bar{\gamma}^{-1} \mathcal{S}_{-}^2 + 240 \bar{\beta}_{-} \bar{\gamma}^{-1} \mathcal{S}_{-} \mathcal{S}_{+}\nn\\
&\qquad\qquad\qquad\qquad\quad  + \delta \Big[80 \bar{\beta}_{-} \mathcal{S}_{-}^2 - 240 \bar{\beta}_{-} \bar{\gamma}^{-1} \mathcal{S}_{-}^2 - 240 \bar{\beta}_{+} \bar{\gamma}^{-1} \mathcal{S}_{-} \mathcal{S}_{+}\Big]  - 80 \mathcal{S}_{-}^2 \nu \Bigg\} \nn\\
& \qquad\qquad\qquad+  4 \pi(2 + \bar{\gamma})  \mathcal{S}_{-}^2 x^{1/2} \nn\\
& \qquad\qquad\qquad + \frac{x}{420} \Bigg\{-2688 \bar{\beta}_{+} \zeta^{-1} + 2910 \zeta^{-1} \bar{\gamma} + 1792 \bar{\beta}_{+} \zeta^{-1} \bar{\gamma} + 896 \zeta^{-1} \bar{\gamma}^2 - 3360 \bar{\beta}_{-}^2 \zeta^{-1} \bar{\gamma}^{-1} - 3360 \bar{\beta}_{+}^2 \zeta^{-1} \bar{\gamma}^{-1}\nn\\
& \qquad\qquad\qquad\qquad\ \ \,\quad  - 7560 \mathcal{S}_{-}^2 - 2240 \bar{\beta}_{-}^2 \mathcal{S}_{-}^2 - 3360 \bar{\beta}_{+} \mathcal{S}_{-}^2 - 2240 \bar{\beta}_{+}^2 \mathcal{S}_{-}^2 - 1680 \bar{\gamma} \mathcal{S}_{-}^2 + 560 \bar{\gamma}^2 \mathcal{S}_{-}^2 - 4480 \bar{\beta}_{-}^2 \bar{\gamma}^{-1} \mathcal{S}_{-}^2 \nn\\
& \qquad\qquad\qquad\qquad\ \ \,\quad - 6720 \bar{\beta}_{+} \bar{\gamma}^{-1} \mathcal{S}_{-}^2 - 4480 \bar{\beta}_{+}^2 \bar{\gamma}^{-1} \mathcal{S}_{-}^2 + 13440 \bar{\beta}_{-}^2 \bar{\gamma}^{-2} \mathcal{S}_{-}^2 + 13440 \bar{\beta}_{+}^2 \bar{\gamma}^{-2} \mathcal{S}_{-}^2 - 2240 \bar{\chi}_{+} \mathcal{S}_{-}^2 \nn\\
& \qquad\qquad\qquad\qquad\ \ \,\quad + 6720 \bar{\gamma}^{-1} \bar{\chi}_{+} \mathcal{S}_{-}^2 + 2240 \zeta \mathcal{S}_{-}^4 + 1120 \zeta \bar{\gamma} \mathcal{S}_{-}^4 - 2240 \bar{\beta}_{-} \mathcal{S}_{-} \mathcal{S}_{+} - 6720 \bar{\beta}_{-} \bar{\gamma}^{-1} \mathcal{S}_{-} \mathcal{S}_{+} \nn\\
& \qquad\qquad\qquad\qquad\ \ \,\quad - 8960 \bar{\beta}_{-} \bar{\beta}_{+} \bar{\gamma}^{-1} \mathcal{S}_{-} \mathcal{S}_{+} + 26880 \bar{\beta}_{-} \bar{\beta}_{+} \bar{\gamma}^{-2} \mathcal{S}_{-} \mathcal{S}_{+} + 6720 \bar{\gamma}^{-1} \bar{\chi}_{-} \mathcal{S}_{-} \mathcal{S}_{+}\nn\\
& \qquad\qquad\qquad\qquad\ \ \,\quad  + \delta \Big[2688 \bar{\beta}_{-} \zeta^{-1} - 1792 \bar{\beta}_{-} \zeta^{-1} \bar{\gamma} + 6720 \bar{\beta}_{-} \bar{\beta}_{+} \zeta^{-1} \bar{\gamma}^{-1} + 3360 \bar{\beta}_{-} \mathcal{S}_{-}^2 + 4480 \bar{\beta}_{-} \bar{\beta}_{+} \mathcal{S}_{-}^2 \nn\\
& \qquad\qquad\qquad\qquad\ \ \,\quad\qquad  + 6720 \bar{\beta}_{-} \bar{\gamma}^{-1} \mathcal{S}_{-}^2 + 8960 \bar{\beta}_{-} \bar{\beta}_{+} \bar{\gamma}^{-1} \mathcal{S}_{-}^2 - 26880 \bar{\beta}_{-} \bar{\beta}_{+} \bar{\gamma}^{-2} \mathcal{S}_{-}^2 + 2240 \bar{\chi}_{-} \mathcal{S}_{-}^2\nn\\
& \qquad\qquad\qquad\qquad\ \ \,\quad\qquad  - 6720 \bar{\gamma}^{-1} \bar{\chi}_{-} \mathcal{S}_{-}^2 + 9240 \mathcal{S}_{-} \mathcal{S}_{+} + 2240 \bar{\beta}_{+} \mathcal{S}_{-} \mathcal{S}_{+} + 5040 \bar{\gamma} \mathcal{S}_{-} \mathcal{S}_{+} + 4480 \bar{\beta}_{-}^2 \bar{\gamma}^{-1} \mathcal{S}_{-} \mathcal{S}_{+} \nn\\
& \qquad\qquad\qquad\qquad\ \ \,\quad\qquad + 6720 \bar{\beta}_{+} \bar{\gamma}^{-1} \mathcal{S}_{-} \mathcal{S}_{+}  + 4480 \bar{\beta}_{+}^2 \bar{\gamma}^{-1} \mathcal{S}_{-} \mathcal{S}_{+} - 13440 \bar{\beta}_{-}^2 \bar{\gamma}^{-2} \mathcal{S}_{-} \mathcal{S}_{+} - 13440 \bar{\beta}_{+}^2 \bar{\gamma}^{-2} \mathcal{S}_{-} \mathcal{S}_{+}\nn\\
& \qquad\qquad\qquad\qquad\ \ \,\quad\qquad  - 6720 \bar{\gamma}^{-1} \bar{\chi}_{+} \mathcal{S}_{-} \mathcal{S}_{+}  + 2240 \zeta \mathcal{S}_{-}^3 \mathcal{S}_{+} + 1120 \zeta \bar{\gamma} \mathcal{S}_{-}^3 \mathcal{S}_{+}\Big]  \nn\\
& \qquad\qquad\qquad\qquad\ \ \,\quad + \nu \Big[1960 \zeta^{-1} \bar{\gamma} + 13440 \bar{\beta}_{+}^2 \zeta^{-1} \bar{\gamma}^{-1} + 11480 \mathcal{S}_{-}^2 + 8960 \bar{\beta}_{-}^2 \mathcal{S}_{-}^2 + 22400 \bar{\beta}_{+} \mathcal{S}_{-}^2 - 1120 \bar{\gamma} \mathcal{S}_{-}^2 \nn\\
& \qquad\qquad\qquad\qquad\ \ \,\quad\qquad+ 44800 \bar{\beta}_{-}^2 \bar{\gamma}^{-1} \mathcal{S}_{-}^2 - 31360 \bar{\beta}_{+} \bar{\gamma}^{-1} \mathcal{S}_{-}^2 - 26880 \bar{\beta}_{+}^2 \bar{\gamma}^{-1} \mathcal{S}_{-}^2 - 80640 \bar{\beta}_{-}^2 \bar{\gamma}^{-2} \mathcal{S}_{-}^2 \nn\\
& \qquad\qquad\qquad\qquad\ \ \,\quad\qquad + 26880 \bar{\beta}_{+}^2 \bar{\gamma}^{-2} \mathcal{S}_{-}^2  + 4480 \bar{\chi}_{+} \mathcal{S}_{-}^2 - 13440 \bar{\gamma}^{-1} \bar{\chi}_{+} \mathcal{S}_{-}^2 - 4480 \zeta \mathcal{S}_{-}^4 - 2240 \zeta \bar{\gamma} \mathcal{S}_{-}^4 \nn\\
& \qquad\qquad\qquad\qquad\ \ \,\quad\qquad - 31360 \bar{\beta}_{-} \bar{\gamma}^{-1} \mathcal{S}_{-} \mathcal{S}_{+}  + 17920 \bar{\beta}_{-} \bar{\beta}_{+} \bar{\gamma}^{-1} \mathcal{S}_{-} \mathcal{S}_{+} \nn\\
& \qquad\qquad\qquad\qquad\ \ \,\quad\qquad - 53760 \bar{\beta}_{-} \bar{\beta}_{+} \bar{\gamma}^{-2} \mathcal{S}_{-} \mathcal{S}_{+} - 13440 \bar{\gamma}^{-1} \bar{\chi}_{-} \mathcal{S}_{-} \mathcal{S}_{+} \Big]  \nn\\& \qquad\qquad\qquad\qquad\ \ \,\quad + \delta \nu   \Big[ -8960 \bar{\beta}_{-} \mathcal{S}_{-}^2 + 11200 \bar{\beta}_{-} \bar{\gamma}^{-1} \mathcal{S}_{-}^2 + 11200 \bar{\beta}_{+} \bar{\gamma}^{-1} \mathcal{S}_{-} \mathcal{S}_{+}\Big]+ 1120 \mathcal{S}_{-}^2 \nu^2  \Bigg\} \nn\\
& \qquad\qquad\qquad+ \frac{\pi x^{3/2}}{30(1 -  \zeta)} \Bigg\{192 \bar{\gamma} - 192 \zeta^{-1} \bar{\gamma} + 96 \bar{\gamma}^2 - 96 \zeta^{-1} \bar{\gamma}^2 - 96 \mathcal{S}_{-}^2 + 160 \bar{\beta}_{+} \mathcal{S}_{-}^2 + 96 \zeta \mathcal{S}_{-}^2 - 160 \bar{\beta}_{+} \zeta \mathcal{S}_{-}^2 - 208 \bar{\gamma} \mathcal{S}_{-}^2  \nn \\
&\qquad\qquad\qquad\qquad\qquad\,\,\qquad - 160 \bar{\beta}_{+} \bar{\gamma} \mathcal{S}_{-}^2 + 208 \zeta \bar{\gamma} \mathcal{S}_{-}^2 + 160 \bar{\beta}_{+} \zeta \bar{\gamma} \mathcal{S}_{-}^2 - 80 \bar{\gamma}^2 \mathcal{S}_{-}^2 + 80 \zeta \bar{\gamma}^2 \mathcal{S}_{-}^2 + 960 \bar{\beta}_{+} \bar{\gamma}^{-1} \mathcal{S}_{-}^2  \nn \\
&\qquad\qquad\qquad\qquad\qquad\,\,\qquad - 960 \bar{\beta}_{+} \zeta \bar{\gamma}^{-1} \mathcal{S}_{-}^2 + 480 \bar{\beta}_{-} \mathcal{S}_{-} \mathcal{S}_{+} - 480 \bar{\beta}_{-} \zeta \mathcal{S}_{-} \mathcal{S}_{+} + 960 \bar{\beta}_{-} \bar{\gamma}^{-1} \mathcal{S}_{-} \mathcal{S}_{+} - 960 \bar{\beta}_{-} \zeta \bar{\gamma}^{-1} \mathcal{S}_{-} \mathcal{S}_{+}  \nn \\
&\qquad\qquad\qquad\qquad\qquad\,\,\qquad + \delta \Big[-160 \bar{\beta}_{-} \mathcal{S}_{-}^2 + 160 \bar{\beta}_{-} \zeta \mathcal{S}_{-}^2 + 160 \bar{\beta}_{-} \bar{\gamma} \mathcal{S}_{-}^2 - 160 \bar{\beta}_{-} \zeta \bar{\gamma} \mathcal{S}_{-}^2 - 960 \bar{\beta}_{-} \bar{\gamma}^{-1} \mathcal{S}_{-}^2 \nn \\
&\qquad\qquad\qquad\qquad\qquad\ \,\qquad\qquad+ 960 \bar{\beta}_{-} \zeta \bar{\gamma}^{-1} \mathcal{S}_{-}^2 - 384 \mathcal{S}_{-} \mathcal{S}_{+} - 480 \bar{\beta}_{+} \mathcal{S}_{-} \mathcal{S}_{+} + 384 \zeta \mathcal{S}_{-} \mathcal{S}_{+} + 480 \bar{\beta}_{+} \zeta \mathcal{S}_{-} \mathcal{S}_{+} \nn \\
&\qquad\qquad\qquad\qquad\qquad\ \,\qquad\qquad- 192 \bar{\gamma} \mathcal{S}_{-} \mathcal{S}_{+} + 192 \zeta \bar{\gamma} \mathcal{S}_{-} \mathcal{S}_{+} - 960 \bar{\beta}_{+} \bar{\gamma}^{-1} \mathcal{S}_{-} \mathcal{S}_{+} + 960 \bar{\beta}_{+} \zeta \bar{\gamma}^{-1} \mathcal{S}_{-} \mathcal{S}_{+}\Big]   \nn \\
&\qquad\qquad\qquad\qquad\qquad\,\,\qquad + \nu \Big[-1208 \mathcal{S}_{-}^2 + 1208 \zeta \mathcal{S}_{-}^2 - 604 \bar{\gamma} \mathcal{S}_{-}^2 + 604 \zeta \bar{\gamma} \mathcal{S}_{-}^2\Big] \Bigg\} \Bigg)\,.
\end{align}
\end{subequations}
Next, we apply the usual flux-balance argument for the total tensor $+$ scalar energy flux,
\begin{align}
\frac{\dd E_\text{circ}}{\dd t} = -\mathcal{F}^\text{total}_\text{circ}\,,\qquad \mathcal{F}^\text{total}_\text{circ} \equiv \mathcal{F}_\text{circ} + \mathcal{F}^s_\text{circ}\,,
\end{align}
where $E_\text{circ}$ denotes the conservative energy of the system deduced from the conservative equations of motion and is given at 2PN order by~\cite{Bernard:2018ivi} 
\begin{align}\label{Ecirc3PN}
E_\text{circ} &= -\frac{1}{2}m\nu c^{2}x\,\Bigg(1
+ \frac{x}{12} \Biggl\{
 \bigl(-9
+ 8 \bar\beta_{+}
- 8 \overline{\gamma}
\bigr) 
- \nu
- 8\bar\beta_{-} \delta
\Biggr\}  \nonumber\\
&\qquad\qquad\qquad\ \quad
+ \frac{x^2}{24}\Biggl\{-81
+ 32 \bar\beta_{+}^2
+ 32 \bar\chi_{+}
+ 8 \bar\delta_{+}
- 112 \overline{\gamma} - 38 \overline{\gamma}^2
+   24\bar\beta_{+}
+ 32 \overline{\gamma} \bar\beta_{+}
+ 32 \bar\beta_{-}^2 \delta^2\nonumber\\
&\qquad\qquad\qquad\qquad\,\qquad 
- \nu^2  + 8  \bar\beta_{-}\nu\delta
+  \nu   \Big[57 -384 \overline{\gamma}^{-1} \bar\beta_{-}^2 + 384 \overline{\gamma}^{-1} \bar\beta_{+}^2 - 152 \bar\beta_{+} 
- 64 \bar\chi_{+}
+ 32 \bar\delta_{+}
+ 88 \overline{\gamma}
+ 8 \overline{\gamma}^2
\Big]\nonumber\\
&\qquad\qquad\qquad\qquad\qquad
+ \delta\Big[ - 24 \bar\beta_{-} 
- 32 \bar\chi_{-}  + 8 \bar\delta_{-} -64 \bar\beta_{+} \bar\beta_{-} - 32 \bar\beta_{-} \overline{\gamma}\Big]
\Biggr\}  
\Biggr)\,.
\end{align}

The orbital phase is defined as a function of time by $\phi_\text{circ} = \int_{t_0}^t \omega(t')\,\dd t'$. Using the flux balance equation, we perform a change of variables to express the phase as a function of $x$,
\be
\phi_\text{circ} = -\frac{c^3}{\tilde{G} \alpha m} \int_{x_0}^x \frac{x'^{3/2}}{\mathcal{F}_\text{circ}(x')} \frac{\dd E_\text{circ}}{\dd x'}\,\dd x'\,.
\ee
Performing the post-Newtonian expansion inside the integral and integrating term by term (following the simplest PN approximant~\cite{Boyle:2008ge}), we find that, up to a constant, the phase is given by
\begin{align}\label{phasecirc}
\phi_\text{circ}& =- \frac{1}{4 \zeta \mathcal{S}_{-}^2 \nu x^{1/2}} \Bigg[ x^{-1}  \nn\\
& \qquad+ \frac{3}{2} + 8 \bar{\beta}_{+} - 2 \bar{\gamma} - 12 \bar{\beta}_{+} \bar{\gamma}^{-1} -  \frac{72}{5} \zeta^{-1} \mathcal{S}_{-}^{-2} - 6 \zeta^{-1} \bar{\gamma} \mathcal{S}_{-}^{-2} - 12 \bar{\beta}_{-} \bar{\gamma}^{-1} \mathcal{S}_{-}^{-1} \mathcal{S}_{+} \nn\\
& \qquad + \delta\Big[-8 \bar{\beta}_{-} + 12 \bar{\beta}_{-} \bar{\gamma}^{-1} + 12 \bar{\beta}_{+} \bar{\gamma}^{-1} \mathcal{S}_{-}^{-1} \mathcal{S}_{+}\Big]  + \frac{7}{2} \nu \nn\\
&\qquad+3 \pi x^{1/2}\log(x)   \left(1 +  \frac{\bar{\gamma}}{2}\right)   \nn\\
&\qquad + x \Bigg\{ \frac{111}{8} -  \frac{80}{3} \bar{\beta}_{-}^2 - 33 \bar{\beta}_{+} -  \frac{80}{3} \bar{\beta}_{+}^2 + \frac{87}{2} \bar{\gamma} -  \frac{44}{3} \bar{\beta}_{+} \bar{\gamma} + \frac{52}{3} \bar{\gamma}^2 + 40 \bar{\beta}_{-}^2 \bar{\gamma}^{-1} + 18 \bar{\beta}_{+} \bar{\gamma}^{-1} + 40 \bar{\beta}_{+}^2 \bar{\gamma}^{-1} - 72 \bar{\beta}_{-}^2 \bar{\gamma}^{-2}\nn\\
&\qquad\qquad\quad - 72 \bar{\beta}_{+}^2 \bar{\gamma}^{-2} - 16 \bar{\chi}_{+} + 12 \bar{\gamma}^{-1} \bar{\chi}_{+} - 2 \zeta \mathcal{S}_{-}^2 -  \zeta \bar{\gamma} \mathcal{S}_{-}^2 -  \frac{1221}{70} \zeta^{-1} \mathcal{S}_{-}^{-2} -  \frac{168}{5} \bar{\beta}_{+} \zeta^{-1} \mathcal{S}_{-}^{-2} -  \frac{1029}{40} \zeta^{-1} \bar{\gamma} \mathcal{S}_{-}^{-2} \nn\\
&\qquad\qquad\quad + 8 \bar{\beta}_{+} \zeta^{-1} \bar{\gamma} \mathcal{S}_{-}^{-2} - 8 \zeta^{-1} \bar{\gamma}^2 \mathcal{S}_{-}^{-2} + 18 \bar{\beta}_{-}^2 \zeta^{-1} \bar{\gamma}^{-1} \mathcal{S}_{-}^{-2} -  \frac{576}{5} \bar{\beta}_{+} \zeta^{-1} \bar{\gamma}^{-1} \mathcal{S}_{-}^{-2} + 18 \bar{\beta}_{+}^2 \zeta^{-1} \bar{\gamma}^{-1} \mathcal{S}_{-}^{-2} \nn\\
&\qquad\qquad\quad -  \frac{1728}{25} \zeta^{-2} \mathcal{S}_{-}^{-4} -  \frac{288}{5} \zeta^{-2} \bar{\gamma} \mathcal{S}_{-}^{-4} - 12 \zeta^{-2} \bar{\gamma}^2 \mathcal{S}_{-}^{-4} - 4 \bar{\beta}_{-} \mathcal{S}_{-}^{-1} \mathcal{S}_{+} + 18 \bar{\beta}_{-} \bar{\gamma}^{-1} \mathcal{S}_{-}^{-1} \mathcal{S}_{+} + 80 \bar{\beta}_{-} \bar{\beta}_{+} \bar{\gamma}^{-1} \mathcal{S}_{-}^{-1} \mathcal{S}_{+} \nn\\
&\qquad\qquad\quad - 144 \bar{\beta}_{-} \bar{\beta}_{+} \bar{\gamma}^{-2} \mathcal{S}_{-}^{-1} \mathcal{S}_{+} + 12 \bar{\gamma}^{-1} \bar{\chi}_{-} \mathcal{S}_{-}^{-1} \mathcal{S}_{+} - 48 \bar{\beta}_{-} \zeta^{-1} \mathcal{S}_{-}^{-3} \mathcal{S}_{+} -  \frac{576}{5} \bar{\beta}_{-} \zeta^{-1} \bar{\gamma}^{-1} \mathcal{S}_{-}^{-3} \mathcal{S}_{+} \nn\\
&\qquad\qquad\quad + \delta \Big[33 \bar{\beta}_{-} + \frac{160}{3} \bar{\beta}_{-} \bar{\beta}_{+} + \frac{44}{3} \bar{\beta}_{-} \bar{\gamma} - 18 \bar{\beta}_{-} \bar{\gamma}^{-1} - 80 \bar{\beta}_{-} \bar{\beta}_{+} \bar{\gamma}^{-1} + 144 \bar{\beta}_{-} \bar{\beta}_{+} \bar{\gamma}^{-2} + 16 \bar{\chi}_{-} - 12 \bar{\gamma}^{-1} \bar{\chi}_{-} \nn\\
&\qquad\qquad\quad\qquad  + \frac{168}{5} \bar{\beta}_{-} \zeta^{-1} \mathcal{S}_{-}^{-2} - 8 \bar{\beta}_{-} \zeta^{-1} \bar{\gamma} \mathcal{S}_{-}^{-2} + \frac{576}{5} \bar{\beta}_{-} \zeta^{-1} \bar{\gamma}^{-1} \mathcal{S}_{-}^{-2} - 36 \bar{\beta}_{-} \bar{\beta}_{+} \zeta^{-1} \bar{\gamma}^{-1} \mathcal{S}_{-}^{-2} - 2 \zeta \mathcal{S}_{-} \mathcal{S}_{+}\nn\\
&\qquad\qquad\quad\qquad  -  \zeta \bar{\gamma} \mathcal{S}_{-} \mathcal{S}_{+} + \frac{33}{2} \mathcal{S}_{-}^{-1} \mathcal{S}_{+} + 4 \bar{\beta}_{+} \mathcal{S}_{-}^{-1} \mathcal{S}_{+} + 9 \bar{\gamma} \mathcal{S}_{-}^{-1} \mathcal{S}_{+} - 40 \bar{\beta}_{-}^2 \bar{\gamma}^{-1} \mathcal{S}_{-}^{-1} \mathcal{S}_{+} - 18 \bar{\beta}_{+} \bar{\gamma}^{-1} \mathcal{S}_{-}^{-1} \mathcal{S}_{+} \nn\\
&\qquad\qquad\quad\qquad - 40 \bar{\beta}_{+}^2 \bar{\gamma}^{-1} \mathcal{S}_{-}^{-1} \mathcal{S}_{+} + 72 \bar{\beta}_{-}^2 \bar{\gamma}^{-2} \mathcal{S}_{-}^{-1} \mathcal{S}_{+} + 72 \bar{\beta}_{+}^2 \bar{\gamma}^{-2} \mathcal{S}_{-}^{-1} \mathcal{S}_{+} - 12 \bar{\gamma}^{-1} \bar{\chi}_{+} \mathcal{S}_{-}^{-1} \mathcal{S}_{+} + 48 \bar{\beta}_{+} \zeta^{-1} \mathcal{S}_{-}^{-3} \mathcal{S}_{+} \nn\\
&\qquad\qquad\quad\qquad + \frac{576}{5} \bar{\beta}_{+} \zeta^{-1} \bar{\gamma}^{-1} \mathcal{S}_{-}^{-3} \mathcal{S}_{+}\Big]   +  \delta \nu \Big[- \frac{11}{3} \bar{\beta}_{-} - 10 \bar{\beta}_{-} \bar{\gamma}^{-1} - 10 \bar{\beta}_{+} \bar{\gamma}^{-1} \mathcal{S}_{-}^{-1} \mathcal{S}_{+}\Big] -  \frac{55}{24} \nu^2  \nn\\
&\qquad\qquad\quad + \nu \Big[- \frac{79}{8} + \frac{320}{3} \bar{\beta}_{-}^2 + \frac{245}{3} \bar{\beta}_{+} -  \frac{86}{3} \bar{\gamma} + 32 \bar{\beta}_{-}^2 \bar{\gamma}^{-1} - 26 \bar{\beta}_{+} \bar{\gamma}^{-1} - 192 \bar{\beta}_{+}^2 \bar{\gamma}^{-1} + 48 \bar{\beta}_{-}^2 \bar{\gamma}^{-2} + 240 \bar{\beta}_{+}^2 \bar{\gamma}^{-2} \nn\\
&\qquad\qquad\quad\qquad  + 32 \bar{\chi}_{+} - 24 \bar{\gamma}^{-1} \bar{\chi}_{+} - 32 \zeta \mathcal{S}_{-}^2 - 16 \zeta \bar{\gamma} \mathcal{S}_{-}^2 - 6 \zeta^{-1} \mathcal{S}_{-}^{-2} -  \frac{5}{2} \zeta^{-1} \bar{\gamma} \mathcal{S}_{-}^{-2} - 72 \bar{\beta}_{+}^2 \zeta^{-1} \bar{\gamma}^{-1} \mathcal{S}_{-}^{-2} \nn\\
&\qquad\qquad\quad\qquad   - 26 \bar{\beta}_{-} \bar{\gamma}^{-1} \mathcal{S}_{-}^{-1} \mathcal{S}_{+} - 160 \bar{\beta}_{-} \bar{\beta}_{+} \bar{\gamma}^{-1} \mathcal{S}_{-}^{-1} \mathcal{S}_{+} + 288 \bar{\beta}_{-} \bar{\beta}_{+} \bar{\gamma}^{-2} \mathcal{S}_{-}^{-1} \mathcal{S}_{+} - 24 \bar{\gamma}^{-1} \bar{\chi}_{-} \mathcal{S}_{-}^{-1} \mathcal{S}_{+} \Big]   \Bigg\} \nn\\
&\qquad  +\frac{\pi x^{3/2}}{1 -  \zeta} \Bigg\{\frac{63}{10} + 2 \bar{\beta}_{+} -  \frac{63}{10} \zeta - 2 \bar{\beta}_{+} \zeta + \frac{23}{20} \bar{\gamma} + 4 \bar{\beta}_{+} \bar{\gamma} -  \frac{23}{20} \zeta \bar{\gamma} - 4 \bar{\beta}_{+} \zeta \bar{\gamma} -  \bar{\gamma}^2 + \zeta \bar{\gamma}^2 - 12 \bar{\beta}_{+} \bar{\gamma}^{-1} + 12 \bar{\beta}_{+} \zeta \bar{\gamma}^{-1}  \nn\\
&\qquad\quad - 6 \bar{\beta}_{-} \mathcal{S}_{-}^{-1} \mathcal{S}_{+} + 6 \bar{\beta}_{-} \zeta \mathcal{S}_{-}^{-1} \mathcal{S}_{+} - 12 \bar{\beta}_{-} \bar{\gamma}^{-1} \mathcal{S}_{-}^{-1} \mathcal{S}_{+} + 12 \bar{\beta}_{-} \zeta \bar{\gamma}^{-1} \mathcal{S}_{-}^{-1} \mathcal{S}_{+}  \nn\\
&\qquad\quad +\delta  \Big[-2 \bar{\beta}_{-} + 2 \bar{\beta}_{-} \zeta - 4 \bar{\beta}_{-} \bar{\gamma} + 4 \bar{\beta}_{-} \zeta \bar{\gamma} + 12 \bar{\beta}_{-} \bar{\gamma}^{-1} - 12 \bar{\beta}_{-} \zeta \bar{\gamma}^{-1} -  \frac{24}{5} \mathcal{S}_{-}^{-1} \mathcal{S}_{+}  + 6 \bar{\beta}_{+} \mathcal{S}_{-}^{-1} \mathcal{S}_{+} \nn\\
&\qquad\qquad\, \quad  + \frac{24}{5} \zeta \mathcal{S}_{-}^{-1} \mathcal{S}_{+} - 6 \bar{\beta}_{+} \zeta \mathcal{S}_{-}^{-1} \mathcal{S}_{+} -  \frac{12}{5} \bar{\gamma} \mathcal{S}_{-}^{-1} \mathcal{S}_{+} + \frac{12}{5} \zeta \bar{\gamma} \mathcal{S}_{-}^{-1} \mathcal{S}_{+} + 12 \bar{\beta}_{+} \bar{\gamma}^{-1} \mathcal{S}_{-}^{-1} \mathcal{S}_{+} - 12 \bar{\beta}_{+} \zeta \bar{\gamma}^{-1} \mathcal{S}_{-}^{-1} \mathcal{S}_{+}\Big]   \nn\\
&\qquad\quad+\nu  \Big[- \frac{38}{5} + \frac{38}{5} \zeta -  \frac{19}{5} \bar{\gamma} + \frac{19}{5} \zeta \bar{\gamma}\Big] \Bigg\}  \Bigg] \,.\end{align}
This expression for the phase is computed using the natural assumption that the dipolar mode is the leading order, \textit{i.e.} using a formal expansion when $x\to 0$. However, Sennett \textit{et al.}~\cite{Sennett:2016klh} pointed out that under certain conditions on the scalar-tensor parameters and on the orbital frequency (or frequency band of a given detector), the quadrupolar mode may be actually dominant over the dipolar mode. This happens when
\be 
1 \lesssim \left( \frac{24}{5 \zeta \mathcal{S}_-^2}\right) \left(\frac{\tilde{G} \alpha m \omega}{2}\right)^{2/3}\, ,
\ee
in which case it is more natural, when computing the phase, to perform the expansion around the quadrupolar term rather that the dipolar term: this is what is called the quadrupole-driven (QD) regime, as opposed to the dipolar-driven (DD) regime which we assumed in~\eqref{phasecirc} (the quadrupolar case can easily be performed if needed). This result for the phase is in perfect agreement with the result found by~\cite{Sennett:2016klh} up to 1PN order in the dipolar-driven case.

\section{Waveform and GW modes}
\label{sec:WF}

We now proceed to compute the spherical harmonic modes in circular orbits, that are useful for numerical relativity. We will perform the standard computation for the tensor modes $h^{\ell m}$ and extend the formalism to the scalar modes~$\psi^{\ell m}$.

For a planar orbit of two particles, the polarization orthonormal triad $(\bm{N},\bm{P},\bm{Q})$ is defined following the conventions of~\cite{Faye:2012we}. The TT tensor $h_{ij}^\text{TT}$ represents the gothic \emph{conformal} metric and is a spin-2 object: it can be, just as in GR, decomposed into two independent modes along the polarization vectors,
\begin{subequations}
\begin{align}
h_+& \equiv \frac{1}{2}(P_i P_j - Q_i Q_j) h_{ij}^{TT}\,,\\
h_\times & \equiv \frac{1}{2}(P_i Q_j + Q_i P_j) h_{ij}^{TT}\,,
\end{align}
\end{subequations}
which can be recast into a complex field, $h = h_+ - \di h_\times$, which can itself be decomposed on the basis on spin-weighted spherical harmonics of weight $-2$,
\begin{subequations}
\be h = h_+ - \di h_- = \sum_{\ell=2}^{+\infty} \sum_{m=-\ell}^\ell h^{\ell m} \,_ {-2}Y^{\ell m}\,.\ee
Similarly, the pure spin-0 scalar field can be decomposed on standard (spin-0) spherical harmonics,
\be \psi =\sum_{\ell=0}^{+\infty} \sum_{m=-\ell}^\ell \psi^{\ell m}\  Y^{\ell m} \,.\ee
\end{subequations}
In an alternative presentation of the waveform~\eqref{GWFa}, we can define the \textquotedblleft electric\textquotedblright and \textquotedblleft magnetic\textquotedblright pure-spin tensor harmonics  based on the spin-weighted spherical harmonics~\cite{Kidder:2007rt},
\begin{subequations}
\begin{align}
	T_{ij}^{E2, \ell m} &= \frac{1}{\sqrt{2}}\Big( \,_{-2}Y^{\ell m} m_i m_j +  \,_{2}Y^{\ell m} m_i^* m_j^* \Big)\,,\\
	T_{ij}^{B2, \ell m} &= \frac{-\di}{\sqrt{2}}\Big( \,_{-2}Y^{\ell m} m_i m_j -  \,_{2}Y^{\ell m} m_i^* m_j^* \Big) \,,
\end{align}
\end{subequations}
where $\bm{m}\equiv (\bm{n} +\di \bm{\lambda})/\sqrt{2}$ and $\bm{m}^*$ denotes its complex conjugate. Owing to the fact that $h_{ij}^\text{TT}$ and $\psi$ are only needed at leading order in the inverse distance of the detector to the source $1/R$, we define the spherical harmonic radiative moments $\mathcal{U}^{\ell m}$, $\mathcal{V}^{\ell m}$\text{ and }$\mathcal{U}_s^{\ell m}$ as the components of the decomposition
\begin{subequations}
\begin{align}
h_{ij}^{\mathrm{TT}}&=\sum_{\ell=2}^{+\infty} \sum_{m=-\ell}^\ell \frac{1}{R}\Bigg(\mathcal{U}^{\ell m} T_{ij}^{\mathrm{E3}, \ell m} + \mathcal{V}^{\ell m} T_{ij}^{\mathrm{B3}, \ell m}\Bigg) + \mathcal{O}\left(\frac{1}{R^2}\right)\,,\\
\psi &=\sum_{\ell=0}^\infty \sum_{m=-\ell}^\ell \frac{1}{R}\mathcal{U}_s^{\ell m} Y^{\ell m} + \mathcal{O}\left(\frac{1}{R^2}\right) \,,
\end{align}
\end{subequations}
which can directly be related to the spin-2 and spin-0 modes via
\begin{subequations}
\begin{align}
\label{hlmUlmVlm}
h^{\ell m} &= - \frac{G}{\sqrt{2} R c^{\ell+2}}\Big[ \mathcal{U}^{\ell m} - \frac{\di}{c} \mathcal{V}^{\ell m}\Big]\,,\\
\psi^{\ell m} &= \frac{G}{R c^{\ell+2}} \mathcal{U}_s^{\ell m}\,.
\end{align}
\end{subequations}
Based on symmetry considerations~\cite{Kidder:2007rt,Faye:2012we}, we can further simplify~\eqref{hlmUlmVlm} in the case when the orbit of the binary system is planar (either no spins or aligned/anti-aligned spins) by noticing that $\mathcal{V}^{\ell m}$ is zero when $\ell+m$ is even and $\mathcal{U}^{\ell m}$ is zero when $\ell+m$ is odd. These spherical harmonic moments are related to the STF multipole moments by
\begin{subequations}
\begin{align}
\mathcal{U}^{\ell m} &= \frac{4}{\ell!}\sqrt{\frac{(\ell+1)(\ell+2)}{2\ell(\ell-1)}}\alpha_L^{\ell m} \mathcal{U}_L\,,\\
\mathcal{V}^{\ell m} &= -\frac{8}{\ell!}\sqrt{\frac{\ell(\ell+2)}{2(\ell+1)(\ell-1)}}\alpha_L^{\ell m} \mathcal{V}_L\,,\\
\mathcal{U}_s^{\ell m} &= -\frac{2}{\ell!} \alpha_L^{\ell m} \mathcal{U}^s_L\,,
\end{align}
\end{subequations}
where we have defined $\alpha_L^{\ell m}$ as~\cite{Henry:2021cek}\footnote{The link with the alternative definition given in~\cite{Thorne:1980ru} is given by $\mathcal{Y}_L^{\ell m}=\frac{(2\ell+1)!!}{4 \pi \ell!}{\left(\alpha_L^{\ell m}\right)}^*$.} 
\be 
\alpha_L^{\ell m} \equiv \int \dd \Omega \,\hat{N}_L {\left(Y^{\ell m}\right)}^* = \frac{\sqrt{4\pi} (-\sqrt{2})^m \ell!}{\sqrt{(2\ell+1)(\ell+m)!(\ell-m)!} }  {m_0^*}^{\langle M} l^{L-M\rangle}\,,
\ee
with $\bm{l}$ the unit normal to the orbit, and $\mathbf{m}^*_0$ the complex conjugate of the vector $\mathbf{m}$ taken at some initial time $t_0$ at which $\mathbf{n}$ is aligned with $\mathbf{P}$~\cite{Faye:2012we}. Finally, it can be shown that the redefinition of the phase variable~\cite{Sennett:2016klh}
\be 
\psi_\text{circ} \equiv \phi_\text{circ} - \frac{2 (1-\zeta)}{\alpha} x^{3/2} \bigg(\log(4 \omega \tau_0)+\gamma_E-\frac{11}{12}\bigg)\,,
\ee
succeeds in removing most logarithms from the expressions of $h^{\ell m}$ and $\psi^{\ell m}$ once the complex exponential is PN-expanded. These modes can then be recast into dimensionless amplitude modes $h^{\ell m}$ and $\psi^{\ell m}$ which are given by\footnote{Recall that $h^{\ell, -m} = (-)^\ell {\left(h^{\ell m}\right)}^*$ and $\psi^{\ell, -m} = (-)^\ell {\left(\psi^{\ell m}\right)}^*$. }
\begin{subequations}
\begin{align}
h^{\ell m} &= \frac{2 \tilde{G}(1-\zeta) m \nu x}{R c^2}  \sqrt{\frac{16\pi}{5}} \,\hat{H}^{\ell m} \de^{-\di m \psi } \,,\\
\psi^{\ell m} &= \frac{2 i \tilde{G} \zeta \sqrt{\alpha}\mathcal{S}_{-} m \nu \sqrt{x}}{R c^2} \sqrt{\frac{8\pi}{3}} \,\hat{\Psi}^{\ell m} \de^{-\di m \psi}\,.
\end{align}
\end{subequations}
with normalized modes defined such that $\hat{H}^{22} = 1 + \mathcal{O}(x)$ and $\hat{\Psi}^{11} = 1 + \mathcal{O}({x})$.

The $\hat{H}^{\ell m}$ were previously computed up to 2PN order by Sennet \textit{et al.}~\cite{Sennett:2016klh}, based on the tensorial waveform of Ref.~\cite{Lang:2013fna}. In this work, we only need these modes to 1.5PN order. We are in complete agreement with their result~\cite{Sennett:2016klh} up to 1.5PN order. In particular the comparison of the mode $\hat{H}^{22}$ at 1.5PN order enables us to check that our formalism for generating the non-linear multipole interactions in~\eqref{radUij} is consistent with the DIRE formalism used by Lang~\cite{Lang:2013fna}.

On the other hand, to our knowledge, the scalar modes $\hat{\Psi}^{\ell m}$ are new to this paper. The expressions for all modes (with non-negative $m$) that are non-zero at 1.5PN order are given by 
\begin{subequations}
\begin{align}
\hat{\Psi}^{00}&=  - \frac{i \sqrt{3} }{\sqrt{2} \mathcal{S}_{-}} \Bigg(\frac{\mathcal{S}_{+} + \mathcal{S}_{-} \delta}{\nu x^{1/2}} + x^{1/2} \Bigg\{8 \bar{\beta}_{-} \bar{\gamma}^{-1} \mathcal{S}_{-} -  \frac{5}{2} \mathcal{S}_{+} + 8 \bar{\beta}_{+} \bar{\gamma}^{-1} \mathcal{S}_{+} + \frac{1}{2} \mathcal{S}_{-} \delta \Bigg\}  \nn\\
&\qquad\quad\qquad  + x^{3/2} \Bigg\{\frac{7}{3} \bar{\beta}_{-} \mathcal{S}_{-} - 4 \bar{\beta}_{-} \bar{\gamma}^{-1} \mathcal{S}_{-} + \frac{16}{3} \bar{\beta}_{-} \bar{\beta}_{+} \bar{\gamma}^{-1} \mathcal{S}_{-} + 4 \bar{\gamma}^{-1} \bar{\chi}_{-} \mathcal{S}_{-} -  \frac{21}{8} \mathcal{S}_{+} + 5 \bar{\beta}_{+} \mathcal{S}_{+} -  \frac{7}{3} \bar{\gamma} \mathcal{S}_{+} - 4 \bar{\beta}_{+} \bar{\gamma}^{-1} \mathcal{S}_{+} \nn\\
&\qquad\quad\qquad\qquad\qquad  + \frac{16}{3} \bar{\beta}_{+}^2 \bar{\gamma}^{-1} \mathcal{S}_{+} - 16 \bar{\beta}_{-}^2 \bar{\gamma}^{-2} \mathcal{S}_{+} + 16 \bar{\beta}_{+}^2 \bar{\gamma}^{-2} \mathcal{S}_{+} + 4 \bar{\gamma}^{-1} \bar{\chi}_{+} \mathcal{S}_{+} \nn\\
&\qquad\quad\qquad\qquad\qquad  + \nu \Big[- \frac{8}{3} \bar{\beta}_{-} \mathcal{S}_{-} + \frac{4}{3} \bar{\beta}_{-} \bar{\gamma}^{-1} \mathcal{S}_{-} -  \frac{7}{24} \mathcal{S}_{+} + \frac{4}{3} \bar{\beta}_{+} \bar{\gamma}^{-1} \mathcal{S}_{+}\Big] + \frac{5}{24} \mathcal{S}_{-} \nu \delta  \nn\\
& \qquad\quad\qquad\qquad\qquad  + \delta \Big[- \frac{3}{8} \mathcal{S}_{-} + \frac{1}{3} \bar{\beta}_{+} \mathcal{S}_{-} -  \frac{1}{3} \bar{\gamma} \mathcal{S}_{-} -  \frac{16}{3} \bar{\beta}_{-}^2 \bar{\gamma}^{-1} \mathcal{S}_{-} - 16 \bar{\beta}_{-}^2 \bar{\gamma}^{-2} \mathcal{S}_{-} + 16 \bar{\beta}_{+}^2 \bar{\gamma}^{-2} \mathcal{S}_{-} \nn\\
&\qquad\quad\qquad\qquad\qquad  \qquad- 4 \bar{\gamma}^{-1} \bar{\chi}_{+} \mathcal{S}_{-} -  \frac{7}{3} \bar{\beta}_{-} \mathcal{S}_{+} -  \frac{16}{3} \bar{\beta}_{-} \bar{\beta}_{+} \bar{\gamma}^{-1} \mathcal{S}_{+} - 4 \bar{\gamma}^{-1} \bar{\chi}_{-} \mathcal{S}_{+} \Big]\Bigg\} \Bigg) \,,\\
\hat{\Psi}^{11}&= 
1 + x \Bigg\{- \frac{9}{5} -  \frac{2}{3} \bar{\beta}_{+} -  \frac{1}{3} \bar{\gamma} + 2 \bar{\beta}_{+} \bar{\gamma}^{-1} + \frac{2 \bar{\beta}_{-} \bar{\gamma}^{-1} \mathcal{S}_{+}}{\mathcal{S}_{-}} + \delta\Big[\frac{2}{3} \bar{\beta}_{-} - 2 \bar{\beta}_{-} \bar{\gamma}^{-1} + \frac{4 \mathcal{S}_{+}}{5 \mathcal{S}_{-}} -  \frac{2 \bar{\beta}_{+} \bar{\gamma}^{-1} \mathcal{S}_{+}}{\mathcal{S}_{-}}\Big]  + \frac{14}{15} \nu\Bigg\} \nn\\
&+ x^{3/2} \Bigg\{\frac{(2+\bar{\gamma})\pi}{2} - i\bigg(\frac{\left(2+\bar{\gamma}\right)\left(1+12 \ln(2)\right) }{12} +  \frac{1}{3} \zeta \left( \mathcal{S}_{+}^2+  \mathcal{S}_{-}^2 \right) 
 + \frac{4}{3}\zeta \mathcal{S}_-^2 \nu
 +  \frac{2}{3} \zeta \mathcal{S}_{-} \mathcal{S}_{+} \delta  \bigg)\Bigg\}   \,, \\
\hat{\Psi}^{22}&= - \frac{i }{\sqrt{5} \mathcal{S}_{-}}  \Bigg( x^{1/2} \Bigg\{- \mathcal{S}_{+} + \mathcal{S}_{-} \delta \Bigg\}  \nn\\
& \qquad\ \, \qquad +x^{3/2} \Bigg\{ \frac{4 \bar{\beta}_{-} \mathcal{S}_{-}}{3} - 4 \bar{\beta}_{-} \bar{\gamma}^{-1} \mathcal{S}_{-} +  \frac{53 \mathcal{S}_{+}}{14 } +  \frac{4 \bar{\beta}_{+} \mathcal{S}_{+}}{3} +  \frac{2 \bar{\gamma} \mathcal{S}_{+}}{3 } - 4 \bar{\beta}_{+} \bar{\gamma}^{-1} \mathcal{S}_{+} \nn\\
& \qquad\ \, \qquad   \qquad\qquad+ \nu\Big[-\frac{16 \bar{\beta}_{-} \mathcal{S}_{-} }{3}   + 8 \bar{\beta}_{-} \bar{\gamma}^{-1} \mathcal{S}_{-} - \frac{211 \mathcal{S}_{+} }{42} +  8 \bar{\beta}_{+} \bar{\gamma}^{-1} \mathcal{S}_{+}\Big] +  \frac{61 \mathcal{S}_{-} \nu\delta }{42}  \nn\\
& \qquad\ \, \qquad   \qquad\qquad+ \delta\Big[ - \frac{53 \mathcal{S}_{-}}{14 } - \frac{4 \bar{\beta}_{+} \mathcal{S}_{-}}{3 } - \frac{2 \bar{\gamma} \mathcal{S}_{-}}{3}  +  4 \bar{\beta}_{+} \bar{\gamma}^{-1} \mathcal{S}_{-}  - \frac{4 \bar{\beta}_{-} \mathcal{S}_{+}}{3} + 4 \bar{\beta}_{-} \bar{\gamma}^{-1} \mathcal{S}_{+}  \Big]  \Bigg\} \Bigg) \,, \\
\hat{\Psi}^{33}&= \frac{9 \sqrt{3} \Big(- \mathcal{S}_{-} + \mathcal{S}_{+} \delta + 2 \mathcal{S}_{-} \nu\Big) x}{4 \sqrt{70} \mathcal{S}_{-}}  \,, \\
\hat{\Psi}^{31}&= - \frac{\Big(- \mathcal{S}_{-} + \mathcal{S}_{+} \delta + 2 \mathcal{S}_{-} \nu\Big) x}{20 \sqrt{14} \mathcal{S}_{-}}  \,,\\
\hat{\Psi}^{44}&= - \frac{16i \Big(\mathcal{S}_{+} -  \mathcal{S}_{-} \delta - 3 \mathcal{S}_{+} \nu + \mathcal{S}_{-} \nu\delta \Big) x^{3/2}}{3 \sqrt{105} \mathcal{S}_{-}}  \,,\\
\hat{\Psi}^{42}&= \frac{2i \Big(\mathcal{S}_{+} -  \mathcal{S}_{-} \delta - 3 \mathcal{S}_{+} \nu + \mathcal{S}_{-} \nu\delta \Big) x^{3/2}}{21 \sqrt{15} \mathcal{S}_{-}}  \,.
\end{align}
\end{subequations}
Note that formally, the dominant scalar mode seems to be $\psi^{00}$. This is misleading because the Newtonian part of $\psi^{00}$ is actually an $x$-independant constant: when taking its derivative to compute the flux, such a term vanishes. The dominant mode is thus indeed $\psi^{11}$, hence our choice of normalization. 



\acknowledgments

The authors warmly thank Sylvain Marsat for his very important help during the verification and comparison of the $\hat{H}^{22}$ mode with his result given in Ref.~\cite{Sennett:2016klh}, which allowed us to spot an error in an initial computation on our side. L. Bernard acknowledges financial support from the ``IdEx Université de Paris, ANR-18-IDEX-0001.''


\appendix


\section{Dissipative PN-odd terms in the equations of motion}
\label{app:EOM}

The conservative equations of motion of compact binaries in ST theory have been obtained up to 3PN order in~\cite{Bernard:2018hta,Bernard:2018ivi}. The PN-even terms $\bm{a}_{A}^\mathrm{N}$, $\bm{a}_{A}^{\mathrm{1PN}}$, $\bm{a}_{A}^{\mathrm{2PN}}$ and $\bm{a}_{A}^{\mathrm{3PN}}$ in the acceleration for arbitrary orbits in a general frame are given by Eqs.~(5.10)--(5-12) of~\cite{Bernard:2018hta}. The dissipative PN-odd terms $\bm{a}_{A}^{\mathrm{1.5PN}}$ and $\bm{a}_{A}^{\mathrm{2.5PN}}$ can be computed based on their expressions in terms of the EW moments as given by~\cite{Mirshekari:2013vb}. We find
\begin{subequations}	
	\begin{align}
	\bm{a}_{1}^{1.5\mathrm{PN}} =&
	\frac{2 \zeta \mathcal{S}_{-}}{3 c^3 r_{12}}\Bigl(\mathcal{S}_{+} + \mathcal{S}_{-}\Bigr) \left(\frac{\tilde{G}\alpha m_1}{r_{12}}\right) \left(\frac{\tilde{G}\alpha m_2}{r_{12}}\right) \Biggl [3 \, (n_{12}v_{12})\, \bm{n}_{12} - \bm{v}_{12}\Biggl]\,,\\
	\bm{a}^{2.5\mathrm{PN}}_1 =& \frac{1}{60 c^5 r_{12}}\left(\frac{\tilde{G}\alpha m_1}{r_{12}}\right) \left(\frac{\tilde{G}\alpha m_2}{r_{12}}\right) \times\nn\\
	&\quad \Bigg(\bm{n}_{12} \Bigg\{\left(\frac{\tilde{G}\alpha m_1}{r_{12}}\right)  (n_{12}v_1)(-288 + 80 \bar{\beta}_{-} - 80 \bar{\beta}_{+} - 80 \bar{\gamma} - 800 \zeta \mathcal{S}_{-}^2 + 320 \bar{\beta}_{-} \zeta \mathcal{S}_{-}^2 - 320 \bar{\beta}_{+} \zeta \mathcal{S}_{-}^2 - 320 \zeta \bar{\gamma} \mathcal{S}_{-}^2\nn\\
	&\qquad\qquad\qquad\qquad\qquad\ \qquad   - 800 \zeta \mathcal{S}_{-} \mathcal{S}_{+}  + 320 \bar{\beta}_{-} \zeta \mathcal{S}_{-} \mathcal{S}_{+} - 320 \bar{\beta}_{+} \zeta \mathcal{S}_{-} \mathcal{S}_{+} - 320 \zeta \bar{\gamma} \mathcal{S}_{-} \mathcal{S}_{+}) \nn\\
	&\qquad\quad + \left(\frac{\tilde{G}\alpha m_2}{r_{12}}\right)  (n_{12}v_1)  (832 + 80 \bar{\beta}_{-} - 1440 \bar{\beta}_{+} + 640 \bar{\gamma} + 1920 \bar{\beta}_{+}^2 \bar{\gamma}^{-1} - 1900 \zeta \mathcal{S}_{-}^2 - 320 \bar{\beta}_{-} \zeta \mathcal{S}_{-}^2 - 320 \bar{\beta}_{+} \zeta \mathcal{S}_{-}^2 \nn\\
	&\qquad\qquad\qquad\qquad\qquad\ \qquad  - 800 \zeta \bar{\gamma} \mathcal{S}_{-}^2 + 960 \bar{\beta}_{-} \zeta \bar{\gamma}^{-1} \mathcal{S}_{-}^2 + 2880 \bar{\beta}_{+} \zeta \bar{\gamma}^{-1} \mathcal{S}_{-}^2 - 3840 \bar{\beta}_{-}^2 \zeta \bar{\gamma}^{-2} \mathcal{S}_{-}^2 - 3840 \bar{\beta}_{+}^2 \zeta \bar{\gamma}^{-2} \mathcal{S}_{-}^2 \nn\\
	&\qquad\qquad\qquad\qquad\qquad\ \qquad  - 740 \zeta \mathcal{S}_{-} \mathcal{S}_{+} - 320 \bar{\beta}_{-} \zeta \mathcal{S}_{-} \mathcal{S}_{+} - 320 \bar{\beta}_{+} \zeta \mathcal{S}_{-} \mathcal{S}_{+} - 320 \zeta \bar{\gamma} \mathcal{S}_{-} \mathcal{S}_{+} + 2880 \bar{\beta}_{-} \zeta \bar{\gamma}^{-1} \mathcal{S}_{-} \mathcal{S}_{+} \nn\\
	&\qquad\qquad\qquad\qquad\qquad\ \qquad  + 960 \bar{\beta}_{+} \zeta \bar{\gamma}^{-1} \mathcal{S}_{-} \mathcal{S}_{+} - 7680 \bar{\beta}_{-} \bar{\beta}_{+} \zeta \bar{\gamma}^{-2} \mathcal{S}_{-} \mathcal{S}_{+})\nn\\
	&\qquad\quad + (n_{12}v_1)^3 (600 \bar{\beta}_{-} - 600 \bar{\beta}_{+} + 300 \bar{\gamma})  \nn\\
	&\qquad\quad + (n_{12}v_1) v_1^2   (144 - 360 \bar{\beta}_{-} + 360 \bar{\beta}_{+} - 120 \bar{\gamma} + 180 \zeta \mathcal{S}_{-}^2 + 120 \zeta \bar{\gamma} \mathcal{S}_{-}^2 + 180 \zeta \mathcal{S}_{-} \mathcal{S}_{+} + 120 \zeta \bar{\gamma} \mathcal{S}_{-} \mathcal{S}_{+}) \nn\\
	&\qquad\quad + \left(\frac{\tilde{G}\alpha m_1}{r_{12}}\right)  (n_{12}v_2) (288 - 80 \bar{\beta}_{-} + 80 \bar{\beta}_{+} + 80 \bar{\gamma} + 880 \zeta \mathcal{S}_{-}^2 - 320 \bar{\beta}_{-} \zeta \mathcal{S}_{-}^2 + 320 \bar{\beta}_{+} \zeta \mathcal{S}_{-}^2 + 320 \zeta \bar{\gamma} \mathcal{S}_{-}^2 \nn\\
	&\qquad\qquad\qquad\qquad\qquad\ \qquad  + 880 \zeta \mathcal{S}_{-} \mathcal{S}_{+} - 320 \bar{\beta}_{-} \zeta \mathcal{S}_{-} \mathcal{S}_{+} + 320 \bar{\beta}_{+} \zeta \mathcal{S}_{-} \mathcal{S}_{+} + 320 \zeta \bar{\gamma} \mathcal{S}_{-} \mathcal{S}_{+}) \nn\\
	&\qquad\quad + \left(\frac{\tilde{G}\alpha m_2}{r_{12}}\right) (n_{12}v_2)   (-832 - 80 \bar{\beta}_{-} + 1440 \bar{\beta}_{+} - 640 \bar{\gamma} - 1920 \bar{\beta}_{+}^2 \bar{\gamma}^{-1} + 1900 \zeta \mathcal{S}_{-}^2 + 320 \bar{\beta}_{-} \zeta \mathcal{S}_{-}^2 + 320 \bar{\beta}_{+} \zeta \mathcal{S}_{-}^2 \nn\\
	&\qquad\qquad\qquad\qquad\qquad\ \qquad  + 800 \zeta \bar{\gamma} \mathcal{S}_{-}^2 - 960 \bar{\beta}_{-} \zeta \bar{\gamma}^{-1} \mathcal{S}_{-}^2 - 2880 \bar{\beta}_{+} \zeta \bar{\gamma}^{-1} \mathcal{S}_{-}^2 + 3840 \bar{\beta}_{-}^2 \zeta \bar{\gamma}^{-2} \mathcal{S}_{-}^2 + 3840 \bar{\beta}_{+}^2 \zeta \bar{\gamma}^{-2} \mathcal{S}_{-}^2 \nn\\
	&\qquad\qquad\qquad\qquad\qquad\ \qquad  + 740 \zeta \mathcal{S}_{-} \mathcal{S}_{+} + 320 \bar{\beta}_{-} \zeta \mathcal{S}_{-} \mathcal{S}_{+} + 320 \bar{\beta}_{+} \zeta \mathcal{S}_{-} \mathcal{S}_{+} + 320 \zeta \bar{\gamma} \mathcal{S}_{-} \mathcal{S}_{+} - 2880 \bar{\beta}_{-} \zeta \bar{\gamma}^{-1} \mathcal{S}_{-} \mathcal{S}_{+}\nn\\
	&\qquad\qquad\qquad\qquad\qquad\ \qquad   - 960 \bar{\beta}_{+} \zeta \bar{\gamma}^{-1} \mathcal{S}_{-} \mathcal{S}_{+} + 7680 \bar{\beta}_{-} \bar{\beta}_{+} \zeta \bar{\gamma}^{-2} \mathcal{S}_{-} \mathcal{S}_{+}) \nn\\
	&\qquad\quad + (n_{12}v_1)^2 (n_{12}v_2) (-1800 \bar{\beta}_{-} + 1800 \bar{\beta}_{+} - 900 \bar{\gamma})  \nn\\
	&\qquad\quad + v_1^2 (n_{12}v_2)  (-144 + 360 \bar{\beta}_{-} - 360 \bar{\beta}_{+} + 120 \bar{\gamma} - 180 \zeta \mathcal{S}_{-}^2 - 120 \zeta \bar{\gamma} \mathcal{S}_{-}^2 - 180 \zeta \mathcal{S}_{-} \mathcal{S}_{+} - 120 \zeta \bar{\gamma} \mathcal{S}_{-} \mathcal{S}_{+}) \nn\\
	&\qquad\quad + (n_{12}v_1) (n_{12}v_2)^2 (1800 \bar{\beta}_{-} - 1800 \bar{\beta}_{+} + 900 \bar{\gamma} - 300 \zeta \mathcal{S}_{-}^2 - 300 \zeta \mathcal{S}_{-} \mathcal{S}_{+})  \nn\\
	&\qquad\quad +(n_{12}v_2)^3   (-600 \bar{\beta}_{-} + 600 \bar{\beta}_{+} - 300 \bar{\gamma} + 300 \zeta \mathcal{S}_{-}^2 + 300 \zeta \mathcal{S}_{-} \mathcal{S}_{+}) \nn\\
	&\qquad\quad + (n_{12}v_1) (v_{1}v_{2})  (-288 + 720 \bar{\beta}_{-} - 720 \bar{\beta}_{+} + 240 \bar{\gamma} - 480 \zeta \mathcal{S}_{-}^2 - 240 \zeta \bar{\gamma} \mathcal{S}_{-}^2 - 480 \zeta \mathcal{S}_{-} \mathcal{S}_{+} - 240 \zeta \bar{\gamma} \mathcal{S}_{-} \mathcal{S}_{+}) \nn\\
	&\qquad\quad + (n_{12}v_2) (v_{1}v_{2}) (288 - 720 \bar{\beta}_{-} + 720 \bar{\beta}_{+} - 240 \bar{\gamma} + 600 \zeta \mathcal{S}_{-}^2 + 240 \zeta \bar{\gamma} \mathcal{S}_{-}^2 + 600 \zeta \mathcal{S}_{-} \mathcal{S}_{+} + 240 \zeta \bar{\gamma} \mathcal{S}_{-} \mathcal{S}_{+})  \nn\\
	&\qquad\quad + (n_{12}v_1) v_2^2(144 - 360 \bar{\beta}_{-} + 360 \bar{\beta}_{+} - 120 \bar{\gamma} + 240 \zeta \mathcal{S}_{-}^2 + 120 \zeta \bar{\gamma} \mathcal{S}_{-}^2 + 240 \zeta \mathcal{S}_{-} \mathcal{S}_{+} + 120 \zeta \bar{\gamma} \mathcal{S}_{-} \mathcal{S}_{+})  \nn\\
	&\qquad\quad +(n_{12}v_2) v_2^2 (-144 + 360 \bar{\beta}_{-} - 360 \bar{\beta}_{+} + 120 \bar{\gamma} - 360 \zeta \mathcal{S}_{-}^2 - 120 \zeta \bar{\gamma} \mathcal{S}_{-}^2 - 360 \zeta \mathcal{S}_{-} \mathcal{S}_{+} - 120 \zeta \bar{\gamma} \mathcal{S}_{-} \mathcal{S}_{+}) \Bigg\} \nn\\
&\quad + \bm{v}_{12} \Bigg\{\left(\frac{\tilde{G}\alpha m_1}{r_{12}}\right) (96 - 80 \bar{\beta}_{-} + 80 \bar{\beta}_{+} + 80 \zeta \mathcal{S}_{-}^2 - 80 \bar{\beta}_{-} \zeta \mathcal{S}_{-}^2  \nn\\
	&\qquad\qquad\qquad\qquad\qquad  + 80 \bar{\beta}_{+} \zeta \mathcal{S}_{-}^2 + 80 \zeta \mathcal{S}_{-} \mathcal{S}_{+} - 80 \bar{\beta}_{-} \zeta \mathcal{S}_{-} \mathcal{S}_{+} + 80 \bar{\beta}_{+} \zeta \mathcal{S}_{-} \mathcal{S}_{+}) \nn\\
	&\qquad\quad  + \left(\frac{\tilde{G}\alpha m_2}{r_{12}}\right) (-384 - 80 \bar{\beta}_{-} + 160 \bar{\beta}_{+} - 240 \bar{\gamma} + 340 \zeta \mathcal{S}_{-}^2 + 80 \bar{\beta}_{-} \zeta \mathcal{S}_{-}^2 + 80 \bar{\beta}_{+} \zeta \mathcal{S}_{-}^2 + 160 \zeta \bar{\gamma} \mathcal{S}_{-}^2  \nn\\
	&\qquad\qquad\qquad\qquad\, \quad - 320 \bar{\beta}_{-} \zeta \bar{\gamma}^{-1} \mathcal{S}_{-}^2 - 320 \bar{\beta}_{+} \zeta \bar{\gamma}^{-1} \mathcal{S}_{-}^2 + 60 \zeta \mathcal{S}_{-} \mathcal{S}_{+} + 80 \bar{\beta}_{-} \zeta \mathcal{S}_{-} \mathcal{S}_{+} + 80 \bar{\beta}_{+} \zeta \mathcal{S}_{-} \mathcal{S}_{+}  \nn\\
	&\qquad\qquad\qquad\qquad\, \quad- 320 \bar{\beta}_{-} \zeta \bar{\gamma}^{-1} \mathcal{S}_{-} \mathcal{S}_{+} - 320 \bar{\beta}_{+} \zeta \bar{\gamma}^{-1} \mathcal{S}_{-} \mathcal{S}_{+})  \nn\\
	&\qquad\quad  + (n_{12}v_1)^2 (-360 \bar{\beta}_{-} + 360 \bar{\beta}_{+} - 180 \bar{\gamma} - 480 \zeta \mathcal{S}_{-}^2 - 240 \zeta \bar{\gamma} \mathcal{S}_{-}^2 - 480 \zeta \mathcal{S}_{-} \mathcal{S}_{+} - 240 \zeta \bar{\gamma} \mathcal{S}_{-} \mathcal{S}_{+})    \nn\\
	&\qquad\quad  + v_1^2(-48 + 120 \bar{\beta}_{-} - 120 \bar{\beta}_{+} + 40 \bar{\gamma} + 100 \zeta \mathcal{S}_{-}^2 + 40 \zeta \bar{\gamma} \mathcal{S}_{-}^2 + 100 \zeta \mathcal{S}_{-} \mathcal{S}_{+} + 40 \zeta \bar{\gamma} \mathcal{S}_{-} \mathcal{S}_{+})   \nn\\
	&\qquad\quad  + (n_{12}v_1) (n_{12}v_2) (720 \bar{\beta}_{-} - 720 \bar{\beta}_{+} + 360 \bar{\gamma} + 840 \zeta \mathcal{S}_{-}^2 + 480 \zeta \bar{\gamma} \mathcal{S}_{-}^2 + 840 \zeta \mathcal{S}_{-} \mathcal{S}_{+} + 480 \zeta \bar{\gamma} \mathcal{S}_{-} \mathcal{S}_{+})   \nn\\
	&\qquad\quad  + (n_{12}v_2)^2 (-360 \bar{\beta}_{-} + 360 \bar{\beta}_{+} - 180 \bar{\gamma} - 300 \zeta \mathcal{S}_{-}^2 - 240 \zeta \bar{\gamma} \mathcal{S}_{-}^2 - 300 \zeta \mathcal{S}_{-} \mathcal{S}_{+} - 240 \zeta \bar{\gamma} \mathcal{S}_{-} \mathcal{S}_{+})   \nn\\
	&\qquad\quad  + (v_{1}v_{2})(96 - 240 \bar{\beta}_{-} + 240 \bar{\beta}_{+} - 80 \bar{\gamma} - 120 \zeta \mathcal{S}_{-}^2 - 80 \zeta \bar{\gamma} \mathcal{S}_{-}^2 - 120 \zeta \mathcal{S}_{-} \mathcal{S}_{+} - 80 \zeta \bar{\gamma} \mathcal{S}_{-} \mathcal{S}_{+})   \nn\\
	&\qquad\quad  + v_2^2  (-48 + 120 \bar{\beta}_{-} - 120 \bar{\beta}_{+} + 40 \bar{\gamma} + 40 \zeta \mathcal{S}_{-}^2 + 40 \zeta \bar{\gamma} \mathcal{S}_{-}^2 + 40 \zeta \mathcal{S}_{-} \mathcal{S}_{+} + 40 \zeta \bar{\gamma} \mathcal{S}_{-} \mathcal{S}_{+}) \Bigg\}\Bigg)\nn\\
	\end{align}	
\end{subequations}
whereas $\bm{a}_{2}^{1.5\mathrm{PN}}$ and $\bm{a}_{2}^{2.5\mathrm{PN}}$ are given by $\switch$. For the relative acceleration in the CM frame, the even terms of the equations of motion were computed in~\cite{Bernard:2018hta}, and the odd terms are given explicitly in Ref.~\cite{Mirshekari:2013vb}.

The expressions of the individual positions $\bm{y}_A$ and velocities $\bm{v}_A$ in the CM frame (i.e. as functions of the CM frame quantities $\bm{n}$ and $\bm{v}$) are only required to 2PN order to compute the flux. The even Newtonian, 1PN and 2PN terms are given in~\cite{Bernard:2018ivi}; the only required odd terms are of order 1.5PN: $\bm{v}_A^{\mathrm{1.5PN}} $ is given in~\cite{Mirshekari:2013vb}, but $\bm{y}_A^{\mathrm{1.5PN}}$ is not. We find
\begin{subequations}
	\begin{align}
	\bm{y}_1^{\mathrm{1.5PN}} &=  - \frac{4 (-1 + \zeta) \zeta \tilde{G} m \mathcal{S}_{-} (\mathcal{S}_{+} + \mathcal{S}_{-} \delta) \nu}{3 c^3 (2 + \bar{\gamma})}\, \bm{v}\,,\\
	\bm{v}_1^{\mathrm{1.5PN}} &= - \frac{8 (-1 + \zeta)^2 \zeta \tilde{G}^2 m^2 \mathcal{S}_{-} (\mathcal{S}_{+} + \mathcal{S}_{-} \delta) \nu}{3 c^3 (2 + \bar{\gamma})^2 r^2} \, \bm{n} \,.
	\end{align}
\end{subequations}

\section{Expressions for the scalar and tensor source moments}
\label{app:moments}

\subsection{The STF moments}

For the flux and waveform computations, the STF moments were first expressed in a general frame, then computed in the center of mass frame where their expression are much simpler. However, there is a notable exception for $I^s_i$, which could not be resolved in the center of mass frame up to 2.5PN order. Indeed, this would require an expression for $\bm{y}_1$ as a function of center-of-mass binary system vectors $\bm{n}$ and $\bm{v}$ up to 2.5PN order. Such an expression would require computing the flux $\mathcal{F}_\mathbf{P}$ of linear momentum $\mathbf{P}$ and the flux $\mathcal{F}_\mathbf{G}$ corresponding to the center-of-mass position $\mathbf{G}$, due to the coupling between the dipole and quadrupole moments. However, for the flux, only the second time derivative of $I^s_i$ was required, whose Newtonian part is proportional to the relative acceleration $\bm{a}\equiv \bm{a}_{12}$, which is already expressed in the center of mass frame; the 1PN and higher order terms then only require knowing the expression for $\bm{y}_1$ to 1.5PN order, so the second time derivative of $I^s_i$ can be fully expressed in the center of mass frame (without the knowledge of $\mathbf{P}$ or $\mathbf{G}$) using  the expression given in~\cite{Bernard:2018ivi}. As for the waveform, the first time derivative of $I^s_i$ is only required to 2PN order, so there is no problem.

Owing to these remarks, we chose not to present the full 2.5PN expression of $I^s_i$ in a general frame, though we have computed its lengthy expression explicitly. The relevant scalar moments in the center-of-mass frame are
\begin{subequations}\label{scalarmoments}

\end{subequations}
The tensor source moments (in addition to the mass type quadrupole tensor moment already given by Eq.~\eqref{Iij}) are
\begin{subequations}\label{tensormoments}
\begin{align}
I_{ijk} &= - \frac{m \nu \delta r^3}{\phi_0} \Bigg[n^{\langle i} n^{j} n^{k\rangle} + \frac{1}{6c^2}\Bigg\{n^{\langle i} n^{j} n^{k\rangle} \left(\frac{\tilde{G}\alpha m}{r}\right) (-5  +13  \nu) + n^{\langle i} n^{j} n^{k\rangle} v^2  (5  -19  \nu) \nn\\
	&\qquad\qquad\qquad\quad +  n^{\langle i} n^{j} v^{k\rangle}  (nv)  (-6  + 12  \nu) + n^{\langle i} v^{j} v^{k\rangle} (6  - 12  \nu) \Bigg\}\Bigg] + \bigO3  \,,\\
	I_{ijkl} &= \frac{m \nu  r^4}{\phi_0}n^{\langle i} n^{j} n^{k} n^{l\rangle}(1 - 3 \nu) +\bigO2 \,,\\
	I_{ijklm} &= - \frac{m \nu\delta  r^5}{\phi_0}  n^{\langle i} n^{j} n^{k} n^{l} n^{m\rangle} (1 - 2  \nu) + \bigO \,,\\
	J_{ij} &= - \frac{m \nu \delta r^2 }{\phi_0} \Bigg[\epsilon^{ab\langle i}  n^{j\rangle}  n^{a}v^{b}  + \frac{1}{28 c^2}\Bigg\{\epsilon^{ab\langle i}n^{j\rangle} n^{a}  v^{b} \left(\frac{\tilde{G}\alpha m}{r}\right) \Big(54 + 42 \bar{\gamma} +60  \nu  \Big) \nn\\
	&\qquad\qquad\qquad\qquad\ \ \qquad+ \epsilon^{ab\langle i}  n^{j\rangle}  n^{a}v^{b}v^2 (13  - 68  \nu) + \epsilon^{ab\langle i} v^{j\rangle} n^{a} v^{b} (nv)  (5  - 10  \nu) \Bigg\} \Bigg] + \bigO3\,,\\
	J_{ijk} &= \frac{m \nu  r^3 }{\phi_0}\epsilon^{ab\langle i} n^{j\rangle} n^{k} n^{a} v^{b}  (1 - 3 \nu) + \bigO2 \,,\\
	J_{ijkl} &= - \frac{m \nu \delta r^4 }{\phi_0}  \epsilon^{ab\langle i} n^{j\rangle} n^{k} n^{a} n^{l}v^{b}  (1 - 2  \nu) +\bigO \,.
\end{align}
\end{subequations} 

The conserved quantities, i.e. the monopole $I$, the conserved ADM mass $M$, the conserved energy $E$ and the angular moment $J_i$ are given to the  relevant PN orders by
\begin{subequations}\label{monopoledipole}
\begin{align}
I &= \frac{M}{\phi_0} = \frac{m}{\phi_0}\Big(1+ \frac{E}{c^2} \Big) \qquad\text{with}\qquad E = \frac{1}{2} m \nu  v^2 - m \nu \left(\frac{\tilde{G}\alpha m}{r}\right) + \bigO \,,\\
J_{i} &=\frac{m \nu  r }{\phi_0}   \epsilon^{iab} n^{a}v^{b} + \bigO \,.
\end{align}
\end{subequations}
Finally the two gauge moments needed in our calculation read
\begin{subequations}\label{gaugeWYi}
\begin{align}
W &= \frac{m \nu r (nv)}{3 \phi_0} + \bigO\,,\\
Y_{i} &= \frac{m \delta \nu r}{10 \phi_0} \Bigg[n^{i} \left(\frac{\tilde{G}\alpha m}{r}\right) + n^{i} v^2 -  3  v^{i} (nv)\Bigg] + \bigO \,.
\end{align}
\end{subequations}

\subsection{Link between STF and EW moments}
\label{sec:EW}

The works~\cite{Lang:2013fna,Lang:2014osa} use the so-called Epstein-Wagoner multipole moments while we are using the STF moments. In the gravitational sector the definitions of the EW moments are the same as in GR. Their relations to the STF moments can be found in Appendix~E of~\cite{Will:1996zj}. In particular the instantaneous part of the tensorial flux to 1.5PN order (discarding the tails) in terms of the EW source moments reads %
\begin{align}\label{fluxEW}
\mathcal{F}_\text{inst} &= \frac{\phi_0}{c^5 G}\Bigg\{ \frac{1}{5}\overset{(3)}{I^{EW}_{ab}}\overset{(3)}{I^{EW}_{ab}}- \frac{1}{15}\overset{(3)}{I^{EW}_{aa}}\overset{(3)}{I^{EW}_{bb}}\nn\\
&\qquad\qquad+\frac{1}{c^2}\Bigg[\frac{11}{105}\overset{(3)}{I_{abc}^{EW}}\overset{(3)}{I_{abc}^{EW}}-\frac{2}{35}\overset{(3)}{I_{abc}^{EW}}\overset{(3)}{I_{acb}^{EW}}+\frac{8}{105}\overset{(3)}{I_{aab}^{EW}}\overset{(3)}{I_{bcc}^{EW}}-\frac{2}{35}\overset{(3)}{I_{aba}^{EW}}\overset{(3)}{I_{bcc}^{EW}}-\frac{1}{21}\overset{(3)}{I_{aac}^{EW}}\overset{(3)}{I_{bbc}^{EW}}+\frac{22}{105}\overset{(3)}{I_{ab}^{EW}}\overset{(3)}{I_{abcc}^{EW}}\nn\\
&\qquad\qquad\qquad\quad - \frac{8}{35}\overset{(3)}{I_{ab}^{EW}}\overset{(3)}{I_{acbc}^{EW}} - \frac{2}{21}\overset{(3)}{I_{aa}^{EW}}\overset{(3)}{I_{bbcc}^{EW}}+\frac{8}{105}\overset{(3)}{I_{aa}^{EW}}\overset{(3)}{I_{bcbc}^{EW}}+\frac{8}{105}\overset{(3)}{I_{ab}^{EW}}\overset{(3)}{I_{ccab}^{EW}}\Bigg] + \bigO4 \Bigg\} \,.
\end{align}
Injecting the expressions of the EW moments \cite{Lang:2013fna} into Eq.~\eqref{fluxEW}, we obtain an expression for the tensorial flux that is in perfect agreement with Eq.~\eqref{tensorflux}.

Concerning the scalar sector, the moments used in Refs.~\cite{Lang:2013fna,Lang:2014osa} are just non-STF moments $I_L^{s,\text{EW}}$ such that the linearized scalar waveform is given by Eq.~(2.17b) of~\cite{Lang:2013fna}. Comparing with our definition of the STF moments $I_L^s\equiv I_L^{s,\text{STF}}$ entering the scalar waveform in~\eqref{psi1}, we obtain the following relation, valid up to any PN order:
\be\label{ILEWSTF}
I_L^s(u) = - \sum_{k=0}^{+\infty} \frac{(2\ell+1)!!}{2^k k!(2\ell+2k+1)!!}\left(\frac{\dd}{c\, \dd u}\right)^{2k} \!I^{s,\text{EW}}_{2K\langle L\rangle}(u)\,.
\ee
With our computations, and taking into account the link between EW and STF moments as well as the result~\eqref{ILEWSTF}, we are in total agreement with the expressions of the source moments found in~\cite{Lang:2013fna,Lang:2014osa}, \textit{i.e.}
\begin{itemize}
	\item $I_{ij}^\mathrm{EW}$,  $I_{ijk}^\mathrm{EW}$, $I_{ijkl}^\mathrm{EW}$, $I_{ijklm}^\mathrm{EW}$, and $I_{ijklmn}^\mathrm{EW}$  in Eq.~(5.10) of~\cite{Lang:2013fna}\,,
	\item $I^{s,\mathrm{EW}}$, $I^{s,\mathrm{EW}}_i$, $I^{s,\mathrm{EW}}_{ij}$, $I^{s,\mathrm{EW}}_{ijk}$, $I^{s,\mathrm{EW}}_{ijkl}$ and $I^{s,\mathrm{EW}}_{ijklm}$ in the CM frame in Eq.~(3.50) of~\cite{Lang:2014osa}\,.
\end{itemize}
However, despite the perfect correspondence and agreement between the results for STF and EW moments, the scalar flux we obtain in the Eqs.~\eqref{Fsdef}--\eqref{Fscoeffs} in Appendix~\ref{app:fluxcoeffs} disagrees at 1PN order with the one published in Ref.~\cite{Lang:2014osa}. This disagreement remains even when we compute the instantaneous 1PN scalar flux directly from the EW moments using the expression (not given in Ref.~\cite{Lang:2014osa})
\begin{align}\label{FscalarEW}
\mathcal{F}_{\mathrm{inst}}^s &= \frac{\phi_0(3+2\omega_0)}{c^3G}\Bigg\{ \frac{1}{3}\overset{(2)}{I^{s,EW}_a}\overset{(2)}{I^{s,EW}_a}\nn\\
&\qquad\qquad\qquad+\frac{1}{c^2}\Bigg[\bigg(\frac{ \overset{(1)}{m_{s1}^{EW}}}{\phi_0 (3+2\omega_0)}\bigg)^2+\frac{1}{60}\overset{(3)}{I^{s,EW}_{aa}}\overset{(3)}{I^{s,EW}_{bb}}+\frac{1}{30}\overset{(3)}{I^{s,EW}_{ab}}\overset{(3)}{I^{s,EW}_{ab}}+\frac{1}{15}\overset{(2)}{I^{s,EW}_a}\overset{(4)}{I^{s,EW}_{abb}}+\frac{1}{3}\overset{(3)}{I_{aa}^{s,EW}}\frac{\overset{(1)}{m_{s1}^{EW}} }{\phi_0(3+2\omega_0)} \Bigg]\nn\\
&\qquad\qquad\qquad+\frac{1}{c^4}\Bigg[\frac{1}{630}\overset{(4)}{I^{s,EW}_{abc}}\overset{(4)}{I^{s,EW}_{abc}}+\frac{1}{420}\overset{(4)}{I^{s,EW}_{aac}}\overset{(4)}{I^{s,EW}_{bbc}}+\frac{1}{60}\overset{(5)}{I^{s,EW}_{aabb}}\frac{\overset{(1)}{m_{s1}^{EW}}}{\phi_0(3+2\omega_0)}+\frac{1}{420}\overset{(2)}{I^{s,EW}_{a}}\overset{(6)}{I^{s,EW}_{abbcc}}\nn\\
&\qquad\qquad\qquad\qquad\qquad+\frac{1}{840}\overset{(3)}{I^{s,EW}_{aa}}\overset{(5)}{I^{s,EW}_{bbcc}}+\frac{1}{210}\overset{(3)}{I^{s,EW}_{ab}}\overset{(5)}{I^{s,EW}_{abcc}}\Bigg]\Bigg\} +\bigO8 \,,
\end{align}
where we pose, following Ref.~\cite{Lang:2014osa} (note that the convention is slightly different from that in Eq.~\eqref{Is}):
\begin{subequations}
\begin{align}
	I^{s,\text{EW}} &= \frac{1}{\phi_0(3+2\omega_0)}\biggl[ m_{s}^{\text{EW}} + \frac{m_{s1}^{\text{EW}}}{c^2} \biggr] \,,\\
	\text{with}\quad
	m^{s,\text{EW}} &\equiv \sum_A m_A (1 - 2 s_A)\,.\end{align}
\end{subequations}

We have also computed the scalar waveform from the EW moments \cite{Lang:2013fna,Lang:2014osa} using the formulae given in Ref.~\cite{Lang:2014osa}, and we retrieve the explicit expression for the 1.5PN scalar waveform given by Ref.~\cite{Lang:2014osa}. Then, when integrating this waveform using Eq.~(6.6) of \cite{Lang:2014osa}, we retrieve the 1PN part of the scalar flux as given by Eqs.~\eqref{Fsdef}-\eqref{Fscoeffs} of Appendix~\ref{app:fluxcoeffs}, but not the 1PN scalar flux as given by \cite{Lang:2014osa}. 

\section{The scalar 1.5PN flux}
\label{app:fluxcoeffs}

As mentioned throughout the article, the scalar flux that we find disagrees at 1PN order with the scalar flux given by Lang in  Ref.~\cite{Lang:2014osa}. The expression for the difference between the two fluxes at 1PN order is
\begin{align}\label{DiffLang}
&\mathcal{F}^s_\text{inst} - \mathcal{F}^{s,\text{Lang}}_\text{inst} =  \frac{m \nu^2}{c^7 r}\left(\frac{\tilde{G}\alpha m}{r}\right)^3  \Bigg\{ -  \frac{32}{3} \bar{\beta}_{+} \zeta \mathcal{S}_{-} \mathcal{S}_{+} \delta v^{4} \nn\\
&\qquad + \left(\frac{\tilde{G}\alpha m}{r}\right)^2 \bigg(\frac{16}{105} \bar{\gamma} + \nu \Big[- \frac{64}{105} \bar{\gamma} + 64 \bar{\beta}_{-}^2 \zeta \bar{\gamma}^{-1} \mathcal{S}_{-}^2 - 64 \bar{\beta}_{+}^2 \zeta \bar{\gamma}^{-1} \mathcal{S}_{-}^2 
- 64 \bar{\beta}_{-}^2 \zeta \bar{\gamma}^{-2} \mathcal{S}_{-}^2 + 64 \bar{\beta}_{+}^2 \zeta \bar{\gamma}^{-2} \mathcal{S}_{-}^2\Big] \bigg) \Bigg\}\,. 
\end{align}

Since the instantaneous part of the scalar flux to 1.5PN order for general orbits in the center-of-mass frame is very long it is convenient to write it in the following form:
\begin{align}\label{Fsdef}\nn
\mathcal{F}^s_\text{inst} &= \frac{1}{3c^3} \frac{m \nu^2}{r}\left(\frac{\tilde{G} \alpha m}{r}\right)^3 \Bigg\{ A^\text{-1PN} + \frac{2}{5c^2}\left(B^\text{N}_1 \frac{\tilde{G} \alpha m}{r} + B^\text{N}_2 (nv)^2+ B^\text{N}_3 v^2 \right)+ \frac{4}{15c^3} C^\text{0.5PN} \frac{\tilde{G} \alpha m}{r}(nv) \\&\qquad\quad+ \frac{1}{140c^4}\left(D^\text{1PN}_1\left(\frac{\tilde{G} \alpha m}{r}\right)^2 + D^\text{1PN}_2 \frac{\tilde{G} \alpha m}{r}(nv)^2 +  D^\text{1PN}_3  \frac{\tilde{G} \alpha m}{r}v^2 + D^\text{1PN}_4 (nv)^4 + D^\text{1PN}_5 (nv)^2 v^2 + D^\text{1PN}_6 v^4\right)\nn\\
&\qquad\quad+  \frac{1}{90c^5}\left(E^\text{1.5PN}_1\left(\frac{\tilde{G} \alpha m}{r}\right)^2(nv)+E^\text{1.5PN}_2 \frac{\tilde{G} \alpha m}{r}(nv)^3 +E^\text{1.5PN}_3\frac{\tilde{G} \alpha m}{r}(nv)v^2\right)\Bigg\}\,,
\end{align}
where the coefficient $A^\text{-1PN}$, following our convention, corresponds to the leading dipolar radiation at $-1$PN order, and the coefficients $B_n^\text{N}$, $C^\text{0.5PN}$, $D_n^\text{1PN}$ and $E_n^\text{1.5PN}$ respectively parametrize the Newtonian, 0.5PN, 1PN and 1.5PN terms. We have obtained
\begin{subequations}\label{Fscoeffs}

\end{subequations}

\bibliography{ListeRef_ST}

\end{document}